\documentclass[10pt, twocolumn, comsoc]{IEEEtran}

\def\argmin{\mathop{\rm arg\,min}}
\usepackage{graphicx,epsfig}
\usepackage[noadjust]{cite}
\usepackage{mcite}
\usepackage{amsfonts,helvet}
\usepackage{fancyhdr}
\usepackage{threeparttable}
\usepackage{epsf,epsfig}
\usepackage{amsthm}
\usepackage{siunitx}
\usepackage{amssymb}
\usepackage{dsfont}
\usepackage{subfigure}
\usepackage{color}
\usepackage{eucal}
\usepackage{amsmath}
\usepackage[ruled,vlined]{algorithm2e}
\usepackage{enumerate}
\usepackage{cancel}
\usepackage{comment}
\usepackage{siunitx}
\usepackage{graphicx}
\usepackage[labelformat=simple]{subcaption}

\usepackage{gensymb}
\usepackage{algpseudocode}
\usepackage{bm}
\usepackage{epstopdf}

\newtheorem{lemma}{Lemma}

\newtheorem{proposition}{Proposition}
\newtheorem{remark}{Remark}

\SetKwInput{KwInput}{Initialize}
\SetKwInput{KwOutput}{Output}
\SetKwProg{Fn}{Stage 1}{}{}
\SetKwProg{Fn}{Stage 2}{}{}

\usepackage{eucal}

\DeclareMathAlphabet{\mathcal}{OMS}{cmsy}{m}{n}

\begin{document}

\title{Integrated Sensing and Communications in Downlink FDD MIMO without CSI Feedback}

\author{Namhyun~Kim, Juntaek~Han, Jinseok~Choi, Ahmed~Alkhateeb, Chan-Byoung Chae, and Jeonghun~Park

\thanks{
This work was supported in part by Institute of Information \& communications Technology Planning \& Evaluation (IITP) grant funded by the Korea government (MSIT) (No. RS-2024-00397216, Development of the Upper-mid Band Extreme massive MIMO (E-MIMO)), and in part by Institute of Information \& communications Technology Planning \& Evaluation (IITP) grant funded by the Korea government (MSIT) (No. RS-2024-00428780, 6G$\cdot$Cloud Research and Education Open Hub).
Namhyun Kim, Juntaek Han, and Jeonghun Park are with the School of Electrical and Electronic Engineering, Yonsei University, Seoul 03722, South Korea (e-mail: {\texttt{namhyun@yonsei.ac.kr; jthan1218@yonsei.ac.kr; jhpark@yonsei.ac.kr}}).
Jinseok Choi is with the School of Electrical Engineering, Korea Advanced Institute of Science and Technology (KAIST), Daejeon 34141, South Korea (e-mail: {\texttt{jinseok@kaist.ac.kr}}).
Ahmed Alkhateeb is with the School of Electrical, Computer and Energy Engineering, Arizona State University, Tempe 85287, AZ, USA (e-mail: {\texttt{alkhateeb@asu.edu}}).
C. B. Chae is with the School of Integrated Technology, Yonsei University, Seoul, 03722, South Korea (e-mail: {\texttt{cbchae@yonsei.ac.kr}}).
}}

\maketitle \setcounter{page}{1} 

\begin{abstract} 
In this paper, we propose a precoding framework for frequency division duplex (FDD) integrated sensing and communication (ISAC) systems with multiple-input multiple-output (MIMO). Specifically, we aim to maximize ergodic sum spectral efficiency (SE) while satisfying a sensing beam pattern constraint defined by the mean squared error (MSE). Our method reconstructs downlink (DL) channel state information (CSI) from uplink (UL) training signals using partial reciprocity, eliminating the need for CSI feedback. 
{\color{black}{To obtain the error covariance matrix of the reconstructed DL CSI, we devise an \textit{observed Fisher information}-based estimation technique. 
Leveraging this, to mitigate interference caused by imperfect DL CSI reconstruction and sensing operations, we propose a rate-splitting multiple access (RSMA) aided precoder optimization method. 
This method jointly updates the precoding vector and Lagrange multipliers by solving the nonlinear eigenvalue problem with eigenvector dependency to maximize SE.}} The numerical results show that the proposed design achieves precise beam pattern control, maximizes SE, and significantly improves the sensing-communication trade-off compared to the state-of-the-art methods in FDD ISAC scenarios.
\end{abstract} 

\section{Introduction}
The convergence of wireless communication and sensing technologies is accelerating the development of intelligent systems such as autonomous driving, smart cities, and environmental monitoring \cite{liu:survery:22, choi:twc:24, liu:tsp:20}. A key enabler of these applications is integrated sensing and communication (ISAC), which allows both functionalities to share infrastructure and spectrum within a unified architecture. ISAC not only improves spectral efficiency but also reduces hardware complexity and energy consumption \cite{liu:survery:22, xu:jstsp:22, wang:twc:24}.

\textcolor{black}{
When deploying ISAC in cellular systems, one of the most critical design choices is the duplexing mode: time division duplex (TDD) or frequency division duplex (FDD) \cite{liu:arxiv:22, liu:survery:22, wang:twc:24}. TDD systems can benefit from channel reciprocity, which simplifies downlink (DL) channel estimation. However, enabling monostatic sensing in TDD typically requires dedicating separate time slots for sensing to avoid collisions between reflected sensing signals and uplink (UL) communication, which is incompatible with current communication protocols \cite{wang:twc:24}. This leads to implementation challenges and may introduce additional latency \cite{liu:arxiv:22}. In contrast, FDD systems allocate separate frequency bands for UL and DL transmissions, enabling simultaneous communication and sensing without interference, as shown in Fig.~1. This allows ISAC functionality \emph{without} modifications to existing communication standards \cite{wang:twc:24, liu:survery:22}. Moreover, FDD is already widely deployed in cellular networks, making it a practical option for near-term ISAC deployment.
}

\textcolor{black}{
To meet the increasing demands of ISAC—such as supporting multiple users, enabling high-resolution sensing, and managing interference—multiple-input multiple-output (MIMO) technology is essential. MIMO provides rich spatial degrees of freedom, allowing advanced beamforming and spatial multiplexing to support both communication and sensing tasks effectively \cite{liu:survery:22, wang:twc:24}. Within this context, transmit precoding plays a crucial role in managing multi-user interference (MUI) and jointly optimizing the performance of both functionalities \cite{choi:twc:24, xu:jstsp:22}. Nonetheless, a fundamental challenge remains: FDD systems lack channel reciprocity, so DL channel state information (CSI) must be obtained through feedback, resulting in non-trivial overhead and potentially degraded performance in dynamic environments \cite{FDD-caire-2023, xu:access:14}. 
}

In this paper, we propose a novel precoding framework for FDD-based massive MIMO-ISAC systems that eliminates the need for CSI feedback by reconstructing DL CSI from UL training signals. Our approach integrates rate-splitting multiple access (RSMA) with error covariance matrix (ECM) estimation to mitigate MUI and also enhance the sensing performance. \textcolor{black}{The proposed framework is particularly suitable for delay-sensitive applications such as smart cities and autonomous driving, where reducing latency and signaling overhead is crucial \cite{liu:arxiv:22}.}

\subsection{Related work}

\textcolor{black}{In the literature on MIMO-ISAC, the design of the transmit precoder has been a key topic.} For example, \cite{liu:tsp:18} developed a branch-and-bound framework to achieve the global optimal solution that minimizes the mean square error (MSE) of the beam pattern under a constant modulus constraint. However, this approach does not assume designated radar symbols, making it sub-optimal compared to {\color{black}{considering respective symbols for communication and sensing}}. To address this, a joint beamforming design for dedicated communication and sensing symbols was proposed in \cite{liu:tsp:20}. The authors in \cite{chen:tsp:21} introduced a generalized Pareto optimization framework, applying a bisection search algorithm to meet the lowest acceptable performance levels for dual functionalities. A recent study \cite{chowdary2024hybrid} proposed a hybrid radar fusion (HRF) scheme for FDD systems, combining monostatic and bistatic sensing to enhance angle-of-arrival (AoA) estimation. \cite{xu:jstsp:22} proposed a RSMA-based precoder optimization technique to mitigate interference caused by added sensing functionality. However, the studies in  \cite{liu:tsp:18, liu:tsp:20, chen:tsp:21, xu:jstsp:22} did not account for the imperfections in CSI, which are inevitable in practical wireless communication systems \cite{caire:tit:10, park:twc:16}.

\begin{figure}[t]
\centering
\includegraphics[width=1\columnwidth]{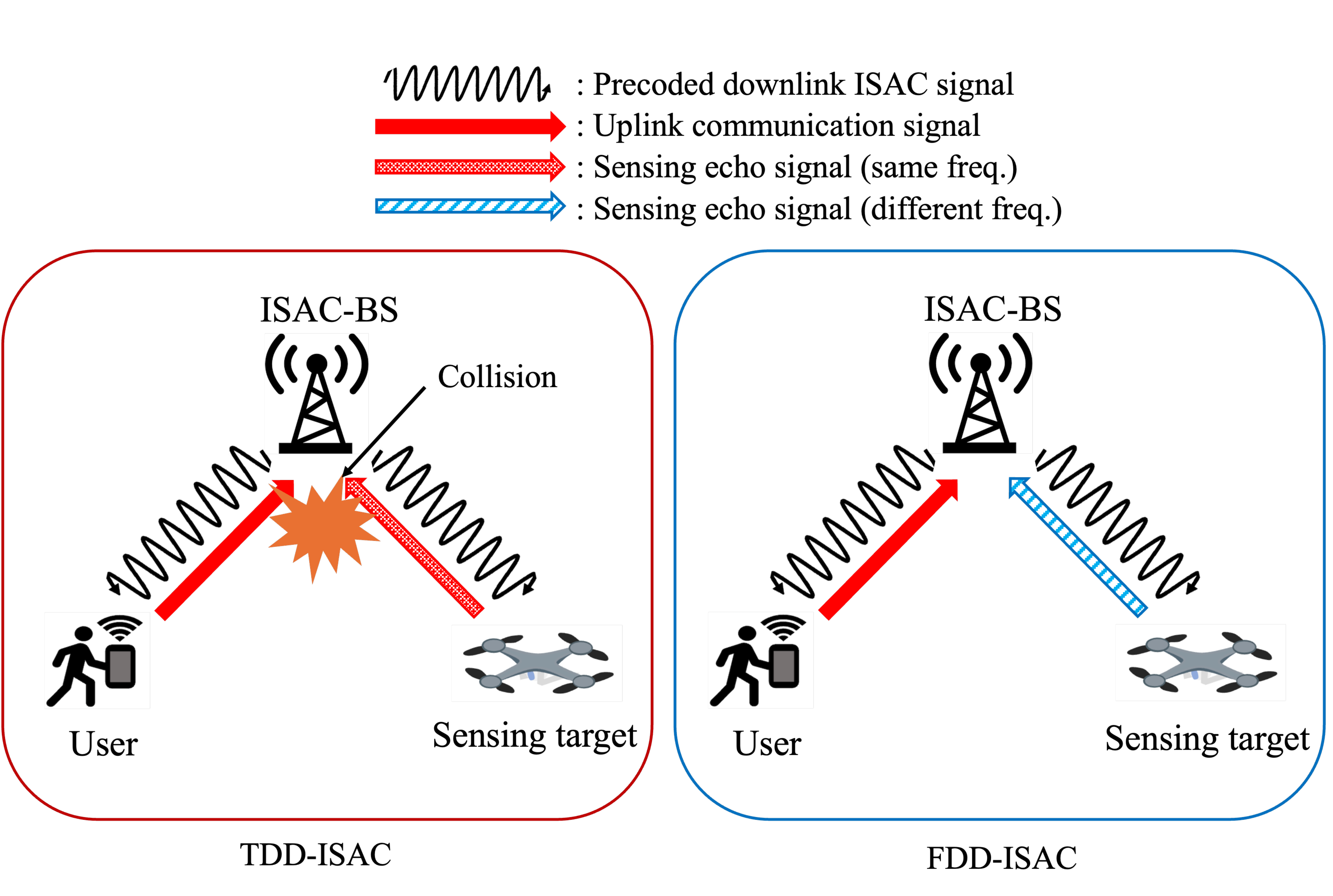}
\caption{\textcolor{black}{TDD ISAC may require modified time slot allocation for signal separation \protect\cite{TD-ISAC}; otherwise, interference between reflected sensing and UL communication signals may occur. In contrast, FDD ISAC leverages separate DL and UL frequency bands to achieve signal separation using established protocols.}}\label{system_model}
\end{figure} 

In light of this, there have also been research endeavors that take into account CSI imperfection. \textcolor{black}{Specifically, in \cite{wang:tsp:24}, channel estimation errors were addressed for MIMO-ISAC, whose key idea is minimizing the worst-case multi-user interference.} In \cite{ren:tcom:22, luan:tits:22}, robust beamforming designs for ISAC were proposed using techniques related to physical layer security or reconfigurable intelligent surface (RIS). 
In addition, robust ISAC beamforming methods for coping with CSI imperfection were also investigated in \cite{bazzi:arxiv:23, ren:tcom:22, luan:tits:22, choi:twc:24}.  
In these works, two types of CSI error modeling were adopted: 
    
    {\textbf{Gaussian error with known covariance }}: The CSI error is modeled as a Gaussian with known covariance \cite{bazzi:arxiv:23, ren:tcom:22, choi:twc:24}. This corresponds to FDD MIMO with scalar quantization.
    
    {\textbf{Bounded error}}: The magnitude of the CSI error is assumed to be bounded within certain ranges \cite{luan:tits:22, ren:tcom:22, wang:tsp:24}. This corresponds to FDD MIMO with vector quantization. 

In summary, the existing error modeling is established under the assumption of channel feedback in FDD MIMO systems.
\textcolor{black}{However, we recall that the amount of feedback often negates SE gains, as it scales with the number of antennas and users \cite{FDD-caire-2023, park:twc:16}, which fundamentally limits the potential gains of MIMO-ISAC; thus developing a low-complexity CSI acquisition approach is essential.}
To address this issue, in \cite{Dina, T.Choi:2021, Han:2019, Rottenberg:2020}, the concept of eliminating CSI feedback was explored in FDD MIMO to simplify the CSI acquisition process, known as zero-feedback strategies. 
A key idea of the zero-feedback strategy is to leverage partial channel reciprocity to efficiently reconstruct the DL channel from UL reference signals, by which we can omit the direct channel feedback even in FDD. 
However, with these DL CSI reconstruction approaches, the CSI error cannot be avoided due to the discrepancies between the UL and DL environments \cite{T.Choi:2021}. 
Unfortunately, it is not feasible to employ the previous CSI error treatment approaches to handle the CSIT error caused by zero-feedback strategies \cite{T.Choi:2021, Rottenberg:2020}. 
\textcolor{black}{That is to say, unlike the approaches in \cite{luan:tits:22, ren:tcom:22, wang:tsp:24}, where CSI quantization ensures a bounded CSI error, the CSIT error in zero-feedback strategies cannot be assumed to be bounded.}
Further, \textit{its error covariance cannot be obtained} in a straightforward way because highly nonlinear algorithms are typically used to reconstruct the DL CSI from UL reference signals such as \cite{NOMP}. 
To resolve this, there have been attempts to approximate the CSI error covariance using the Cramér-Rao lower bound (CRLB) matrix \cite{kim:arxiv:24, Rottenberg:2020}. Specifically, in \cite{kim:arxiv:24}, they computed the CRLB matrix in an instantaneous manner during UL training. Incorporating this into RSMA precoder design, significant SE gains were achieved in FDD MIMO systems without requiring CSI feedback. These findings highlight the potential of a zero-feedback strategy for MIMO-ISAC systems.

\subsection{Contributions}
In this paper, we present a novel FDD MIMO-ISAC transmission framework that does not necessitate CSI feedback. 
Unlike a conventional approach, we reconstruct the DL CSI solely based on UL sounding reference signals by following \cite{Han:2019,kim:arxiv:24}. 
In this setup, acquiring perfect DL CSI is infeasible, making efficient mitigation of interference arising from imperfect DL CSI crucial to achieve robust SE and ensure sensing performance. In this perspective, we summarize our main contributions as follows.
\begin{itemize}
    \item We investigate a DL channel reconstruction method in which channel parameters are estimated solely from UL training. The main problem is then formulated to maximize the sum SE under imperfect DL CSI while meeting a beam pattern matching constraint based on a target MSE. To address CSI inaccuracies caused by zero feedback, we consider the RSMA technique. The communication rate for the common message is defined using a minimum function to ensure that it is decodable by all users. For the sensing constraint, we represent the MSE between the desired and actual beam pattern in a quadratic form. In solving the formulated problem, we find that accurate CSI error covariance matrix estimation is crucial to designing precoders that effectively cope with CSI inaccuracies.
    \item To obtain the error covariance matrix, we propose a novel method, assuming partial reciprocity in the given channel model. In particular, we note that the MSE of the UL channel estimated with the 2D-NOMP estimator closely approaches the CRLB level as demonstrated in \cite{Han:2019}. Based on this, we estimate the error covariance matrix by evaluating the CRLB using the UL received signal. Specifically, we exploit the observed Fisher information, which effectively estimates the realized error when only the \textit{estimated} channel parameter is available. We highlight that the proposed method can effectively provide robustness in the absence of direct CSI feedback.
    \item To solve the formulated problem, we provide an efficient precoding method inspired by the KKT (Karush-Kuhn-Tucker) conditions, which alternately optimizes the precoding vector and the Lagrange multiplier to satisfy the KKT conditions without the need for any off-the-shelf optimization tools. Specifically, we use the principle of nonlinear eigenvalue problem with eigenvector dependency (NEPv) \cite{Park:2023} for designing the precoding vectors. 
    We also find that the Lagrange multiplier dictates the design orientation between sensing and communication, which implies the versatility of the proposed method.
\end{itemize}
Our approach is particularly advantageous in that it achieves robust SE and also precise sensing performance in FDD MIMO systems, all while eliminating the need for CSI feedback. Through this, we provide valuable design insights for FDD MIMO-ISAC  systems with low overhead. 

\textit{Notation}:
The superscripts $(\cdot)^{\sf T}$, $(\cdot)^{\sf H}$, $(\cdot)^{-1}$, $(\cdot)^{\dagger}$ denote the transpose, Hermitian, matrix inversion, and Moore-Penrose pseudo-inverse, respectively. ${\bf{I}}_N$ is the identity matrix of size $N \times N$, Assuming that ${\bf{A}}_1, ..., {\bf{A}}_N \in \mathbb{C}^{K \times K}$, ${\bf{A}} = {\rm blkdiag}\left({\bf{A}}_1, ...,{\bf{A}}_n,..., {\bf{A}}_N \right)$ is a block-diagonal matrix concatenating ${\bf{A}}_1, ..., {\bf{A}}_N$. The Frobenius norm is denoted by $\|\mathbf{A}\|_F$. $\mathbf{A}\circ\mathbf{B}$ denotes the Hadamard product of the two matrices. $\Re(\cdot)$ and $\Im(\cdot)$ denote the real and imaginary parts of a complex quantity, respectively. We use $\lfloor\cdot\rfloor, \lceil\cdot\rceil$ to denote the rounding of a decimal number to its nearest lower and higher integers, respectively. $[K]$ is the set of natural numbers less than or equal to $K$.
\section{System Model}

In this section, we describe the UL and DL channel model and introduce the ISAC signal model assisted by RSMA. 


\subsection{Communication channel model}
Consider a MIMO system with orthogonal frequency division multiplexing (OFDM), in which a base station (BS) has a uniform linear array (ULA) consisting of $N$ antennas and each user is equipped with a single antenna. We assume that there are a total of $S$ pilots for the UL band, with their frequency spacing denoted as $\Delta f$, which implicitly represents the bandwidth of each coherence block. The total bandwidth is set as $B = S \Delta f$.
Adopting the multipath model used in \cite{Han:2019, Dina}, we represent the narrowband baseband UL channel from user $k$ to $n$-th BS antenna over the $s$-th sub-carrier ($k \in [K], n\in [N]$ and $s \in [S]$) as follows:
\begin{align}
    h_{k,n,s}^{\mathrm{ul}} = \sum \limits^{L_k^{\mathrm{ul}}}_{\ell=1}\mathrm{\alpha}^{\mathrm{ul}}_{k, \ell}e^{-j2\pi (n-1)\phi^{\mathrm{ul}}_{k,
    \ell,s}}e^{-j2\mathrm{\pi}(s-\lceil\frac{S}{2}\rceil-1)\Delta f \tau^{\mathrm{ul}}_{k, \ell}},\label{eq:ul channel}
\end{align}
assuming half wavelength array spacing, $\phi_{k,\ell,s}^{\mathrm{ul}}$ is given by
\begin{align}
    \phi_{k,\ell,s}^{\mathrm{ul}} = \frac{\sin\theta_{k, \ell}}{2}\left( 1+\frac{s\Delta f}{f_{\mathrm{c}}^{\mathrm{ul}}} \right), \label{eq:phi_theta}
\end{align} and $\alpha^{\mathrm{ul}}_{k, \ell}\in \mathbb{C}^{1\times 1}$, $\tau^{\mathrm{ul}}_{k, \ell}\in\mathbb{R}^{1\times 1}$ denote the complex path gain and propagation delay of the $\ell$-th path for user $k$ respectively. Both are assumed to be consistent in entire bandwidth for UL \cite{Deokhwan, Han:2019}. We assume that the propagation delay is given by $0<\tau^{\mathrm{ul}}_{k, \ell}<1/\Delta f$, and the number of paths as $L_k^{\mathrm{ul}}$. The AoA of $\ell$-th path for user $k$ is denoted as $\theta^{\mathrm{ul}}_{k, \ell}$, the wavelength of UL carrier frequency as $\lambda_{\mathrm{c}}^{\mathrm{ul}}$, and the antenna spacing for UL reception as $d^{\mathrm{ul}}$ with $d^{\mathrm{ul}} = \lambda_{\mathrm{c}}^{\mathrm{ul}}/2 $. Similarly, the DL channel from $n$-th BS antenna to user $k$ over the $s$-th sub-carrier of DL channel is expressed as
\begin{align}
     h_{k,n,s}^{\mathrm{dl}}(f) = \sum \limits^{L_k^{\mathrm{dl}}}_{\ell=1}\mathrm{\alpha}^{\mathrm{dl}}_{k, \ell}e^{-j2\pi (n-1)\phi^{\mathrm{dl}}_{k,
    \ell,s}}e^{-j2\mathrm{\pi}(f+(s-\lceil\frac{S}{2}\rceil-1)\Delta f)\tau^{\mathrm{dl}}_{k, \ell}},\label{eq:dl channel}
\end{align}
in which we assume that $f$ denotes the frequency difference between carrier frequency of UL and DL, and the DL parameters such as $L_k^{\mathrm{dl}}, \mathrm{\alpha}^{\mathrm{dl}}_{k, \ell}, \phi^{\mathrm{dl}}_{k,\ell,s}, \theta_{k, \ell}^{\mathrm{dl}}, \tau_{k, \ell}^{\mathrm{dl}}$ is defined in the same manner as UL. As reported in numerous previous studies with actual measurements \cite{Dina, Han:2019}, the UL and DL channels share the following parameters:
\begin{itemize}
    \item \textbf{Propagation delay}: $\tau_{k,\ell}^{\mathrm{ul}} = \tau_{k,\ell}^{\mathrm{dl}}$.
    \item \textbf{Angular parameter}: $\theta_{k,\ell}^{\mathrm{ul}} = \theta_{k,\ell}^{\mathrm{dl}}$.
    \item \textbf{Channel path number}: $L_k^{\mathrm{ul}} = L_k^{\mathrm{dl}} \triangleq L_k$.
\end{itemize}

However, channel gains may not exhibit perfect reciprocity \cite{Zhong:2020, Han:2019}. To account for this, we model the DL channel gain as
\begin{align}\label{gain-update}
\alpha_{k, \ell}^{\mathrm{dl}} = \eta_{k,\ell}\alpha_{k, \ell}^{\mathrm{ul}} + \sqrt{1-\eta_{k,\ell}^2}\beta_{k, \ell},
\end{align}
where $\beta_{k, \ell}$ is assumed to be a random variable following $\mathcal{CN}(0, \sigma^2_{\mathrm{path}, k}), \forall \ell$, independent of the UL channel gain $\alpha_{k,\ell}^{\mathrm{ul}}$. \textcolor{black}{We consider a more generalized channel model compared to previous works~\cite{Dina, Deokhwan}, which assumed perfectly reciprocal channel gains and fixed angular parameters. In contrast, our model allows non-reciprocal channel gains ($\eta_{k,\ell} \neq 1$) and accommodates variations in angular characteristics across UL and DL, thus capturing practical deviations in FDD systems more accurately.}

\subsection{RSMA-assisted MIMO-ISAC signal model}
\textcolor{black}{In this study, we notice that the unified ISAC architecture may restrict the spatial channel resources between the user and BS. To effectively address this, we consider RSMA, a multiple access technique that enhances robustness when spatial resources are shared for multiple objectives \cite{RSMA-ten-promising}. Specifically, RSMA mitigates interference between sensing and communication, as well as among communication users \cite{xu:jstsp:22, WMMSE-SAA, Park:2023}. In RSMA, each user's message \(M_k\) is split into a common part \(M_{\mathrm{c}, k}\) and a private part \(M_{\mathrm{p}, k}\). The common parts are encoded into a single common stream \(s_{\mathrm{c}}\), while private parts are encoded individually as \(s_k\), \(k \in [K]\). At the receiver, each user \(k\) first decodes the common stream, removes it using successive interference cancellation (SIC), and then decodes the private stream \(s_k\) while treating residual interference as noise \cite{RSMA-ten-promising}.}

To fully exploit the spatial degrees of freedom (DoF) available in the considered MIMO system, the BS uses dedicated radar signals combined with communication symbols for transmission. Accordingly, the transmitted signal, denoted as ${\bf{x}} \in \mathbb{C}^{N }$, is given as
\begin{align}
    \mathbf{x} &= \mathbf{p}_{\mathrm{c}}s_{\mathrm{c}}+\mathbf{P}_{\mathrm{p}}\mathbf{s}_{\mathrm{p}}+\mathbf{P}_{\mathrm{r}}\mathbf{s}_{\mathrm{r}} \nonumber \\
    &= [\mathbf{P}_{\mathrm{c}}, \mathbf{P}_{\mathrm{r}}][\mathbf{s}_{\mathrm{c}}^{\sf T}, \mathbf{s}_{\mathrm{r}}^{\sf T}]^{\sf T} = \mathbf{P}\mathbf{s},
\end{align}
where $\mathbf{p}_\mathrm{c}$ denotes the precoding vector for the common message symbol $s_\mathrm{c}$, $\mathbf{P}_\mathrm{p}$ is a precoding matrix for private message vector $\mathbf{s}_\mathrm{p} = [s_{\mathrm{p}, 1},s_{\mathrm{p}, 2},\ldots,s_{\mathrm{p}, K}]^{\sf T}$, which is defined as $\mathbf{P}_{\mathrm{p}} = [\mathbf{p}_{\mathrm{p}, 1}, \mathbf{p}_{\mathrm{p}, 2}, \ldots,\mathbf{p}_{\mathrm{p}, K}] \in \mathbb{C}^{N \times K}, $ where $\mathbf{p}_{\mathrm{p}, k}$ is the precoder for the private message of user $k$. Likewise, $\mathbf{P}_\mathrm{r}$ represents the precoding matrix for the vector of radar sequence symbols $\mathbf{s}_\mathrm{r}=[s_{\mathrm{r}, 1},s_{\mathrm{r}, 2},\ldots,s_{\mathrm{r}, M}]^{\sf T}$, that is, $\mathbf{P}_\mathrm{r} = [\mathbf{p}_{\mathrm{r}, 1}, \mathbf{p}_{\mathrm{r}, 2},\ldots,\mathbf{p}_{\mathrm{r}, M}] \in \mathbb{C}^{N \times M}$, where $\mathbf{p}_{\mathrm{p}, m}$ is the radar beamformer for the $m$-th radar symbol. For notational simplicity, we integrate the common and private part with a single expression, which is given by $\mathbf{P}_{\mathrm{c}} = [\mathbf{p}_{\mathrm{c}}, \mathbf{P}_\mathrm{p}]$ and $\mathbf{s}_{\mathrm{c}} = [s_{\mathrm{c}}, \mathbf{s}_\mathrm{p}^{\sf T}]^{\sf T}$. Lastly, we define the entire precoding vector and the symbol vector as $\mathbf{P} = [\mathbf{P}_\mathrm{c}, \mathbf{P}_\mathrm{r}]$ and $\mathbf{s} = [\mathbf{s}_\mathrm{c}^{\sf T}, \mathbf{s}_\mathrm{r}^{\sf T}]^{\sf T}$, respectively. \textcolor{black}{The communication symbols $\mathbf{s}_\mathrm{c}$ and the radar sequences $\mathbf{s}_\mathrm{r}$ are statistically independent, as established in our previous work \cite{choi:twc:24}}. That is, $\mathbb{E}[\mathbf{s}_\mathrm{c}\mathbf{s}_\mathrm{r}^{\sf H}] = \mathbf{0}_{(K+1)\times M}$. In addition, the power constraint with transmit power $P$ is represented by $\mathbb{E}[{\mathbf{s}_\mathrm{c}\mathbf{s}_\mathrm{c}^{\sf H}}] = P \cdot \mathbf{I}_{K+1}$ and $\mathbb{E}[{\mathbf{s}_\mathrm{r}\mathbf{s}_\mathrm{r}^{\sf H}}] = P \cdot \mathbf{I}_M$, respectively. In this regard, we presume that $\|\mathbf{P}_\mathrm{c}\|_{\sf F} + \|\mathbf{P}_\mathrm{r}\|_{\sf F} \leq 1$, without loss of generality.

\section{Downlink channel reconstruction \& Performance Metrics}
In this section, we explore a DL channel reconstruction method based on UL training and define performance metrics for communication and sensing, respectively.

\subsection{Reconstructing downlink channel}

We start by introducing the vector \(\mathbf{u}(\tau_{k,\ell}, \theta_{k,\ell}) \in \mathbb{C}^{NS \times 1}\) which encapsulates the spatial signature along with the delay profile for each sub-carrier. The specific element corresponding to the \(n\)-th antenna and the \(s\)-th sub-carrier is 
\begin{align}
[{\mathbf{u}}(\tau_{k, \ell},\theta_{k,\ell})]_{n, s} = e^{-j2\pi(n-1)\phi^{\mathrm{ul}}_{k,\ell,s}}e^{-j2\mathrm{\pi}(s-\lceil\frac{S}{2}\rceil-1)\Delta f \tau^{\mathrm{ul}}_{k, \ell}},\label{u_comp}
\end{align}
which comes from the channel model in \eqref{eq:ul channel}. 

Assuming that the UL sounding reference signal comprises ones for all sub-carriers and antennas without loss of generality, the received signal vector \(\mathbf{y}_k \in \mathbb{C}^{NS \times 1}\) that represents the signals across all antennas and sub-carriers is described by 
\begin{align}
\mathbf{y}_k = \sum_{\ell=1}^{L_k} \alpha^{\mathrm{ul}}_{k,\ell} \mathbf{u}(\tau^{\mathrm{ul}}_{k,\ell}, \theta^{\mathrm{ul}}_{k,\ell}) + \mathbf{n}_k,\label{y_k original}
\end{align}
where \(\mathbf{n}_k\in \mathbb{C}^{NS\times 1}\) follows additive white Gaussian noise (AWGN). Our aim is to estimate the DL channel from the UL-observed signal \(\mathbf{y}_k\). To achieve this, we first identify the set of parameters of the UL channel \(\left\{{\hat{\alpha}}^{\mathrm{ul}}_{k,\ell}, {\hat{\tau}}^{\mathrm{ul}}_{k,\ell}, \hat{\theta}^{\mathrm{ul}}_{k,\ell}\right\}_{\ell=1,\dots,L_k}\) from \(\mathbf{y}_k\), then reconstruct the DL channel at the frequency difference $f$ by exploiting their frequency invariance discussed in Section II.

For this task, we employ the 2D-NOMP algorithm, which is a state-of-the-art compressed sensing method \cite{NOMP, Han:2019}. This approach is particularly useful, as it circumvents the need for knowledge of number of channel paths while operating efficiently in a grid-less context. Its effectiveness has been validated in practical settings, as demonstrated in \cite{Han:2019}. The detailed procedure of the 2D-NOMP algorithm is elaborated in \cite{Han:2019, kim:arxiv:24}; and we omit it in this paper to conserve space.

As the output, we obtain the quadruplets of estimated UL parameters with $\hat{\alpha}^{\mathrm{ul}}_{k,\ell} = \Re\{\hat{\alpha}^{\mathrm{ul}}_{k,\ell}\}+j\Im\{\hat{\alpha}^{\mathrm{ul}}_{k,\ell}\}$ i.e.,
\begin{align}
\hat{\mathcal{P}}_k = \left\{\Re\{\hat{\alpha}^{\mathrm{ul}}_{k,\ell}\}, \Im\{\hat{\alpha}^{\mathrm{ul}}_{k,\ell}\}, \hat{\tau}^{\mathrm{ul}}_{k,\ell}, \hat{\theta}^{\mathrm{ul}}_{k,\ell}\right\}_{\ell=1, \dots, L_k}, \label{est_para}\end{align} they are used to estimate the DL channel parameters \(\left\{\hat{\alpha}^{\mathrm{dl}}_{k,\ell}, \hat{\tau}^{\mathrm{dl}}_{k,\ell}, \hat{\theta}^{\mathrm{dl}}_{k,\ell}\right\}_{\ell=1, \dots, L_k}\), considering the frequency invariance property. Taking the randomness of channel gain in \eqref{gain-update} into account, we determine the estimated channel parameter for DL as
\begin{align}
    \left\{\hat{\alpha}^{\mathrm{dl}}_{k,\ell}, \hat{\tau}^{\mathrm{dl}}_{k,\ell}, \hat{\theta}^{\mathrm{dl}}_{k,\ell}\right\} \triangleq \left\{\eta_{k, \ell}\hat{\alpha}^{\mathrm{ul}}_{k,\ell}, \hat{\tau}^{\mathrm{ul}}_{k,\ell}, \hat{\theta}^{\mathrm{ul}}_{k,\ell}\right\}, \forall (k, \ell).
\end{align}
With this, we can reconstruct the DL channel in \eqref{eq:dl channel}, which is represented by 
\begin{align}
    \hat{h}_{k,n,s}^{\mathrm{dl}}(f) = \sum \limits^{L_k^{\mathrm{dl}}}_{\ell=1}\hat{\mathrm{\alpha}}^{\mathrm{dl}}_{k, \ell}e^{-j2\pi (n-1)\hat{\phi}^{\mathrm{dl}}_{k,
    \ell,s}}e^{-j2\mathrm{\pi}(f+(s-\lceil\frac{S}{2}\rceil-1)\Delta f)\hat{\tau}^{\mathrm{dl}}_{k, \ell}},\label{eq:dl channel_est}
\end{align} where $\hat{\phi}_{k, \ell, s}^{\mathrm{ul}}$ is defined as
\begin{align}
    \hat{\phi}_{k,\ell,s}^{\mathrm{ul}} = \frac{\sin\hat{\theta}_{k, \ell}}{2}\left( 1+\frac{s\Delta f}{f_{\mathrm{c}}^{\mathrm{ul}}} \right). \label{eq:phi_theta_est}
\end{align}
Correspondingly, the spatial reconstructed channel of (narrowband) $s$-th resource block at $f$ for user $k$ is defined as 
\begin{align}
    \hat{\mathbf{h}}_{k,s}(f) = [\hat{h}_{k,1,s}^{\mathrm{dl}}(f), \hat{h}_{k,2,s}^{\mathrm{dl}}(f), \ldots,  \hat{h}_{k,N,s}^{\mathrm{dl}}(f)] \in \mathbb{C}^{N\times 1}.
\end{align}
For notational simplicity, our signal model and performance metrics that we will find later omit the subcarrier index $s$ and the extrapolation range $f$, assuming that our interested narrowband DL channel is fixed. Thus, \(\hat{\mathbf{h}}_{k,s}(f)\) and its ground truth \({\mathbf{h}}_{k,s}(f)\) will be simply referred to as \(\hat{\mathbf{h}}_k\) and \({\mathbf{h}}_k\), respectively.

\textcolor{black}{\begin{remark}(On the assumption of partial reciprocity): \label{remark_partial}
\normalfont The assumption of reciprocal delay, angular, and path parameters between UL and DL channels (i.e., partial reciprocity) is valid primarily under moderate frequency separations (e.g., several hundred MHz to low GHz), as supported by prior works~\cite{Han:2019, Dina}. 
Especially, we consider a small UL-DL frequency separation (\( f \approx 190 \; {\text{MHz}} \ll f_c^{\mathrm{ul}}, f_c^{\mathrm{dl}} \)), which results in minimal angular spread differences (less than $0.5$ degrees), as presented in 3GPP TR 38.901. 
For significantly larger separations (e.g., multiple GHz), the CSI reconstruction accuracy may degrade due to increased angular parameter mismatch \cite{shakya2024urban}. 
In such cases, we can employ a Gaussian approximation technique to account for angular mismatches, enabling analytical tractability and robust performance evaluation (i.e., the generalized mutual information framework \cite{GantiLapidoth2000}).
\end{remark}}

\subsection{Communication \& sensing performance metric}
In this subsection, we aim to characterize the communication performance metric given the imperfect CSI, and also to define the sensing performance metric for the design of the beam pattern. 

\subsubsection{Communication performance metric} We consider the ergodic sum SE as the communication performance metric. To investigate this, we first represent the received signal at user $k$, denoted as $y_k$, as follows:
\begin{equation}
    y_k = {\mathbf{h}}_{k}^{\sf H} {\mathbf{p}}_{\mathrm{c}} s_{\mathrm{c}}
    +\sum _{i = 1}^{K} {\mathbf{h}}_{k}^{\sf H} {\mathbf{p}}_{\mathrm{p}, i} s_{\mathrm{p}, i}+\sum _{m = 1}^{M} {\mathbf{h}}_{k}^{\sf H} {\mathbf{p}}_{\mathrm{r}, m} s_{\mathrm{r}, m}+ z_{k}, \label{eq:rx_signal}
\end{equation}
where $z_{k} \sim \mathcal {CN}(0,\sigma ^{2})$ follows AWGN.
With this signal model, we aim to maximize the ergodic sum SE under imperfect CSI. Given that the channel estimate $\hat{\mathbf{h}}_k$ is provided, the instantaneous SE for the common and private messages is defined as the expectation over the CSI error $\mathbf{e}_k$:
\textcolor{black}{
\begin{align}
    &R_{\mathrm{c}, k}^{\mathrm{i}}(\mathbf{P})\notag\\ &= \mathbb{E}_{\{\mathbf{e}_k\}}\left[\left.\log_2\left(1+\frac{|\mathbf{h}_k^{\sf H}\mathbf{p}_{\mathrm{c}}|^2}{\begin{Bmatrix} 
\sum_{i=1}^{K}|\mathbf{h}^{\sf H}_k\mathbf{p}_{\mathrm{p}, i}|^2+\\\sum_{m=1}^{K}|\mathbf{h}^{\sf H}_k\mathbf{p}_{\mathrm{r}, m}|^2+\sigma^2/P
    \end{Bmatrix}
    }\right)\right|\hat{\mathbf{h}}_k\right],\label{eq:R_c}\\
    &R_{\mathrm{p}, k}^{\mathrm{i}}(\mathbf{P})\notag\\ &= \mathbb{E}_{\{\mathbf{e}_k\}}\left[\left.\log_2\left(1+\frac{|\mathbf{h}_k^{\sf H}\mathbf{p}_{\mathrm{p}, k}|^2}{
    \begin{Bmatrix}
    \sum_{i=1, i\neq k}^{K}|\mathbf{h}^{\sf H}_k\mathbf{p}_{\mathrm{p}, i}|^2+\\
    \sum_{m=1}^{K}|\mathbf{h}^{\sf H}_k\mathbf{p}_{\mathrm{r}, m}|^2+\sigma^2/P    
    \end{Bmatrix}
    }\right)\right|\hat{\mathbf{h}}_k\right]\label{eq:R_k}.
\end{align}}
{\color{black}{We confirm that the common symbol is not included in the interference term of \eqref{eq:R_k} as it has already been decoded by the RSMA process. This RSMA decoding order (decoding the common message first, followed by the private message) is predetermined, which is a key feature to simplify decoder structure while achieving interference mitigation gains. For more details, we refer to \cite{RSMA-ten-promising}.
}}

The instantaneous rate of the common message $s_{\mathrm{c}}$ is set to ensure decodability for all users, i.e.,
\begin{align}
    R_{\mathrm{c}}^{\mathrm{i}}(\mathbf{P}) = \min_{k\in[K]}R_{\mathrm{c}, k}^{\mathrm{i}}(\mathbf{P}).\label{common-rate-original}
\end{align}
Note, however, that the expressions in \eqref{eq:R_c}, \eqref{eq:R_k} are not in closed form as they are represented with the expectation. To use them as an objective function for our problem, it is required to develop this into a tractable representation. Accordingly, using the direct relationship $\mathbf{h}_k = \hat{\mathbf{h}}_k + \mathbf{e}_k$, the received signal in \eqref{eq:rx_signal} is represented by
\begin{align}
    y_k = &\hat{\mathbf{h}}_{k}^{\sf H} {\mathbf{p}}_{\mathrm{c}} s_{\mathrm{c}}
    +\sum _{i = 1}^{K} \hat{\mathbf{h}}_{k}^{\sf H} {\mathbf{p}}_{\mathrm{p}, i} s_{\mathrm{p}, i}+\sum _{m = 1}^{M} \hat{\mathbf{h}}_{k}^{\sf H} {\mathbf{p}}_{\mathrm{r}, m} s_{\mathrm{r}, m} \notag \\
    &+\underbrace{{\mathbf{e}}_{k}^{\sf H} {\mathbf{p}}_{\mathrm{c}} s_{\mathrm{c}}
    +\sum _{i = 1}^{K} {\mathbf{e}}_{k}^{\sf H} {\mathbf{p}}_{\mathrm{p}, i} s_{\mathrm{p}, i}+\sum _{m = 1}^{M} {\mathbf{e}}_{k}^{\sf H} {\mathbf{p}}_{\mathrm{r}, m} s_{\mathrm{r}, m}}_{(a)}+ z_{k}.
\end{align}
Treating each term in (a) as independent Gaussian noise, the average received signal power at user \( k \), normalized by \( P \) given the channel state, is denoted as follows:
\begin{equation}
\begin{aligned}
&\frac{\mathbb{E}\big\{|y_{k}|^{2}\big\}}{P} \\
&=\overbrace{|\hat{\mathbf{h}}_{k}^{\sf{H}}\mathbf{p}_{\mathrm{c}}|^{2}}^{S_{\mathrm{c}, k}} 
+\underbrace{\overbrace{|\hat{\mathbf{h}}_{k}^{\sf{H}}\mathbf{p}_{\mathrm{p}, k}|^{2}}^{S_{k}} 
+\overbrace{\sum_{i=1, i\neq k}^K |\hat{\mathbf{h}}_{k}^{\sf{H}}\mathbf{p}_{\mathrm{p}, i}|^{2}+\sum_{m=1}^M |\hat{\mathbf{h}}_{k}^{\sf{H}}\mathbf{p}_{\mathrm{r}, m}|^{2}}^{I_k}}_{I_{\mathrm{c}, k}}\\
&\qquad +\overbrace{\underbrace{{|\mathbf{e}}_{k}^{\sf H} {\mathbf{p}}_{\mathrm{c}}|^2
+\sum _{i = 1}^{K} {|\mathbf{e}}_{k}^{\sf H} {\mathbf{p}}_{\mathrm{p}, i}|^2+\sum _{m = 1}^{M} {|\mathbf{e}}_{k}^{\sf H} {\mathbf{p}}_{\mathrm{r}, m}|^2+\frac{\sigma^2}{P}}_{I_{\mathrm{c, k}}}}^{I_k},
\label{eq:original_power}
\end{aligned}
\end{equation}
where we indicate the power of the desired signal in the common part as $S_{\mathrm{c}, k}$ and in the private part as $S_k$, and the corresponding interference as $I_{\mathrm{c}, k}$ and $I_k$, respectively. 
In fact, the independent Gaussian noise assumption suggests the worst case of mutual information to derive a lower bound of \eqref{eq:R_c}, \eqref{eq:R_k} \cite{Park:2023}. Specifically, a lower bound of instantaneous SE for the common message is further derived as
\begin{align}
    &R_{\mathrm{c}, k}^{\mathrm{i}}(\mathbf{P}) \geq \mathbb{E}_{\{\mathbf{e}_k\}}\left[\log_2(1+S_{\mathrm{c}, k}I^{-1}_{\mathrm{c}, k})\right] \\
    &\textcolor{black}{\mathop {\ge}^{(a)}\log _{2} \left ({1 + \frac {|\hat {\mathbf{h}}_{k}^{\sf H} {\mathbf{p}}_{\mathrm{c}}|^{2}}{\begin{Bmatrix} {\sum _{i = 1}^{K} |\hat {\mathbf{h}}_{k}^{\sf H} {\mathbf{p}}_{\mathrm{p}, i} |^{2} +\sum _{m = 1}^{M} |\hat {\mathbf{h}}_{k}^{\sf H} {\mathbf{p}}_{\mathrm{r}, m} |^{2}}
\\ {+{\mathbf{p}}_{\mathrm{c}}^{\sf H} \mathbb {E} \left [{{\mathbf{e}}_{k} {\mathbf{e}}_{k}^{\sf H} }\right] {\mathbf{p}}_{\mathrm{c}}+\sum _{i=1}^{K} {\mathbf{p}}_{\mathrm{p}, i}^{\sf H} \mathbb {E} \left [{{\mathbf{e}}_{k} {\mathbf{e}}_{k}^{\sf H} }\right] {\mathbf{p}}_{\mathrm{p}, i}}\\
+{\sum _{m=1}^{M} {\mathbf{p}}_{\mathrm{r}, m}^{\sf H} \mathbb {E} \left [{{\mathbf{e}}_{k} {\mathbf{e}}_{k}^{\sf H} }\right] {\mathbf{p}}_{\mathrm{r}, m}}+{\sigma ^{2}}/{P}
\end{Bmatrix} } }\right)} \label{R_c_up}\\
&\textcolor{black}{\mathop {\ge}^{(b)}\log _{2} \left ({1 + \frac {|\hat {\mathbf{h}}_{k}^{\sf H} {\mathbf{p}}_{\mathrm{c}}|^{2}}{\begin{Bmatrix} {\sum _{i = 1}^{K} |\hat {\mathbf{h}}_{k}^{\sf H} {\mathbf{p}}_{\mathrm{p}, i} |^{2} +\sum _{m = 1}^{M} |\hat {\mathbf{h}}_{k}^{\sf H} {\mathbf{p}}_{\mathrm{r}, m} |^{2}}
\\ {+{\mathbf{p}}_{\mathrm{c}}^{\sf H} {\boldsymbol{\Sigma }}_{k} {\mathbf{p}}_{\mathrm{c}}+\sum _{i=1}^{K} {\mathbf{p}}_{\mathrm{p}, i}^{\sf H} {\boldsymbol{\Sigma }}_{k} {\mathbf{p}}_{\mathrm{p}, i}}\\
+{\sum _{m=1}^{M} {\mathbf{p}}_{\mathrm{r}, m}^{\sf H} {\boldsymbol{\Sigma }}_{k} {\mathbf{p}}_{\mathrm{r}, m}}+ {\sigma ^{2}}/{P}
\end{Bmatrix} } }\right)}  \label{R_c_k} \\
&\triangleq \bar{R}_{\mathrm{c}, k}^{\mathrm{i}}({\mathbf{P}}) \nonumber.
\end{align}
We note that the inequality (a) follows from the Jensen's inequality, i.e., $\mathbb{E}[\log_2(1+{1}/{x})]\geq\log_2(1+1/\mathbb{E}[x])$, where we omit the subscript of ${\mathbf{e}_k}$ due to the space limitation. $(b)$ comes from $\mathbb {E}[{\bf{e}}_k {\bf{e}}_k^{\sf H}] = {\boldsymbol{\Sigma}}_{k}$. Similar to \eqref{common-rate-original}, we can define a lower bound for the instantaneous rate of common message as well, i.e.,
\begin{align}
    \bar{R}_{\mathrm{c}}^{\mathrm{i}}(\mathbf{P}) = \min_{k\in[K]} \bar{R}^{\mathrm{i}}_{\mathrm{c}, k}(\mathbf{P}). \label{R_c}
\end{align}
Recall that our objective is to maximize the ergodic sum SE. Considering multiple fading processes, the ergodic SE of the common message is determined as
\begin{align}
    R_{\mathrm{c}} = \min_{k\in \mathcal{K}}\left\{\mathbb{E}_{\{\hat{\mathbf{h}}_k\}}\left[R_{\mathrm{c}, k}^{\mathrm{i}}(\mathbf{P})\right]\right\}&\geq\min_{k\in \mathcal{K}}\left\{\mathbb{E}_{\{\hat{\mathbf{h}}_k\}}\left[\bar{R}_{\mathrm{c}, k}^{\mathrm{i}}(\mathbf{P})\right]\right\}\notag\\
    &\geq\mathbb{E}_{\{\hat{\mathbf{h}}_k\}}\left[\bar{R}_{\mathrm{c}}^{\mathrm{i}}(\mathbf{P})\right\}.\label{ergodic_common}
\end{align}
Notably, we will use \eqref{ergodic_common} as our objective function to maximize the ergodic rate of the common message, which corresponds to maximizing the lower bound in \eqref{R_c} for each fading block.
Similarly, a lower bound for the instantaneous SE of the private message is represented as
\begin{align}
&R_{k}^{\mathrm{i}}(\mathbf{P}) \geq \mathbb{E}_{\{\mathbf{e}_k\}}\left[\log_2(1+S_k{I^{-1}_k})\right],\\
&\textcolor{black}{\mathop {\ge }^{}\log _{2} \!\left ({\!1 \!+ \!\frac {|\hat {\mathbf{h}}_{k}^{\sf H} {\mathbf{p}}_{\mathrm{p}, k}|^{2}} {\begin{Bmatrix} {\sum _{i = 1, i \neq k}^{K} |\hat {\mathbf{h}}_{k}^{\sf H} {\mathbf{p}}_{\mathrm{p}, i} |^{2}} +{\sum _{m = 1}^{M} |\hat {\mathbf{h}}_{k}^{\sf H} {\mathbf{p}}_{\mathrm{r}, m} |^{2}+{\mathbf{p}}_{\mathrm{c}}^{\sf H} {\boldsymbol{\Sigma }}_{k} {\mathbf{p}}_{\mathrm{c}}}\\ {+ \sum _{i = 1}^{K} {\mathbf{p}}_{\mathrm{p}, i}^{\sf H} {\boldsymbol{\Sigma }}_{k} {\mathbf{p}}_{\mathrm{p}, i}+ \sum _{m = 1}^{M} {\mathbf{p}}_{\mathrm{r}, m}^{\sf H} {\boldsymbol{\Sigma }}_{k} {\mathbf{p}}_{\mathrm{r}, m} + \frac {\sigma ^{2}}{P}}\end{Bmatrix} }\!}\right)} \label{R_k} \\
&\triangleq \bar{R}^{\mathrm{i}}_{k}({\mathbf{{P}}}),
\end{align}
which is also connected to its ergodic SE as
\begin{align}
    R_k = \mathbb{E}_{\{\hat{\mathbf{h}}_k\}}\left[R_{k}^{\mathrm{i}}(\mathbf{P})\right]\geq\mathbb{E}_{\{\hat{\mathbf{h}}_k\}}\left[\bar{R}_{k}^{\mathrm{i}}(\mathbf{P})\right].\label{ergodic_private}
\end{align}
From \eqref{ergodic_common} and \eqref{ergodic_private}, we derive the following lower bound on the ergodic sum SE, i.e., ${R}_{\sum}:=R_\mathrm{c}+\sum_{k=1}^KR_k$. That is,
\begin{align}
    {R}_{\sum}\geq\bar{R}_{\sum} = \mathbb{E}_{\{\hat{\mathbf{h}}_k\}}\left[\bar{R}^{\mathrm{i}}_{\mathrm{c}}(\mathbf{P})+\sum_{k=1}^K\bar{R}^{\mathrm{i}}_k(\mathbf{P})\right].\label{R_sum}
\end{align}
To maximize ${R}_{\sum}$, we use $\bar{R}_{\sum}$ as a surrogate objective. 

\subsubsection{Sensing performance metric}
As a sensing performance metric, we consider the MSE between the target beam pattern and the actual beam pattern. It should be noted that the MSE metric offers versatility, enabling features such as adaptive sensing beam widening to address angle uncertainties \cite{yin-mse:tcl:22, choi:twc:24, liu:tsp:20}.
\textcolor{black}{Comparisons of MSE and other possible sensing metrics such as CRLB \cite{fan-crb:tsp:22} and SCNR (signal-to-clutter and noise ratio) are presented in our previous work \cite{choi:twc:24}.}

Meanwhile, notice that we have two types of symbols, $\mathbf{s}_{\mathrm{c}}$ and $\mathbf{s}_{\mathrm{r}}$, both of which are precoded to be transmitted as $\mathbf{x}$. Then we define the baseband narrowband signal in the direction of $\theta$ as
\begin{align} q(\theta) = \mathbf{a}^{\sf H}(\theta)\mathbf{x}. \end{align}
Using this, the average signal power for radar toward the direction $\theta$ is represented by
\begin{align}
    G(\mathbf{P};\theta) = \mathbb{E}[|q(\theta)|^2]= P \cdot \mathbf{a}^{\sf H}(\theta)\mathbf{P}\mathbf{P}^{\sf H}\mathbf{a}(\theta).
\end{align}
We then compare this with the normalized target beam pattern $t(\theta_{u})$ to define the sensing beam pattern MSE.
Specifically, assuming \(L\) grids, where the detection angle $\theta_{u}$ is evenly sampled into \(L\) points, the MSE is defined as 
\begin{equation}
    \mathrm{MSE}_{\mathrm{r}} = \frac{1}{L}\sum_{u = 1}^{L} |G(\mathbf{P};\theta_u)-P\cdot t(\theta_u)|^2.\label{MSE_metric}
\end{equation}
With this, we can also easily design sensing beams with multiple target angles by simply letting the target beam pattern for the desired direction as 1, i.e., $t(\theta_u)=1$ otherwise as 0, i.e., $t(\theta_u)=0$. 

\textcolor{black}{\begin{remark}(On the choice of sensing metric)
\normalfont
While radar-specific sensing metrics such as the CRLB and the ambiguity function offer theoretical insights into estimation accuracy and resolution, they may be less limited in certain aspects. Specifically, the CRLB requires precise knowledge of the target parameters or their distributions, which are typically unavailable during initial sensing \cite{fan-crb:tsp:22}. In contrast, the beam-pattern MSE enables explicit control over angular power distribution without relying on target-specific information \cite{choi:twc:24, xu:jstsp:22}. Accordingly, it is thus well-suited for beam sweeping and adaptive sensing under uncertainty. For these reasons, beam-pattern MSE is adopted as the primary sensing metric in this work. Later, we also show that the proposed method can be extended to incorporate other radar performance metrics such as SCNR.
\end{remark}}

\subsection{Problem formulation}
Note that maximizing the ergodic sum SE under the target radar beam pattern MSE $T_{\mathrm{mse}}$ is our main design objective. To achieve this, we aim to maximize the instantaneous achievable SE for each fading block \cite{WMMSE-SAA} and formulate this as $\mathscr{P}_0$:
\begin{align}
    \mathscr{P}_0:& \mathop \textrm{maximize}_{{\mathbf{P}}} \ \min_{k\in[K]} {R}_{\mathrm{c}, k}^{\mathrm{i}}(\mathbf{P})  + \sum_{k=1}^K R^{\mathrm{i}}_{k}
(\mathbf{P}) \label{P0}\\
    &\textrm{subject to } \mathrm{MSE}_{\mathrm{r}} \leq T_{\text{mse}},\label{P0_MSE}\\ &\qquad \ \qquad\|\mathbf{P}\|_{\sf F}^2\leq1.\label{P0_TX_Power}
\end{align}
As previously discussed, with the power constraint on the transmitted symbols $\mathbf{s} = [\mathbf{s}_\mathrm{c}^{\sf T}, \mathbf{s}_\mathrm{r}^{\sf T}]^{\sf T}$ given by the budget \(P\), the precoder norm is constrained to 1. 


\subsection{Problem reformulation}

For the problem $\mathscr{P}_0$, we notice that the expression in \eqref{P0} is not a tractable form. To address this, we use a lower bound derived in the previous section, characterized in \eqref{R_c}, \eqref{R_sum} to maximize the instantaneous SE for each fading block. In addition, considering that the sum SE increases with transmit power \cite{Park:2023}, we just assume $\|\mathbf{P}\|_{\sf F}^2=1$. Consequently, the reformulated problem is as follows:
\begin{align}\label{P1}
    \mathscr{P}_1:& \mathop \textrm{maximize}_{\mathbf{P}} \min_{k\in [K]} \bar{R}^{\mathrm{i}}_{\mathrm{c}, k}({\mathbf{P}}) + \sum_{k=1}^{K} \bar{R}^{\mathrm{i}}_k({\mathbf{P}}) \\
    &\textrm{subject to } \textrm{MSE}_{\textrm{r}} \leq T_{\text{mse}}, \\
    &\qquad \ \qquad \|\mathbf{P}\|^2_{\sf F}=1.\label{P1_constraint}
\end{align}

Now, we rewrite the precoding matrix in vector form, that is, $\bar{\mathbf{p}} = \mathrm{vec}(\mathbf{P}) \in \mathbb{C}^{N(K+M+1)},$ and $\|\bar{\mathbf{p}}\|^2=1$.
By replacing $\mathbf{P}$ with $\bar{\mathbf{p}}$, the SINR terms for the common message of user $k$ in \eqref{R_c_k} and the private message of user $k$ in \eqref{R_k} are reformulated into a tractable form, denoted as $\gamma_{\mathrm{c}}(k)$ and $\gamma_k$, respectively, as follows:
 \begin{align}
     \gamma_{\mathrm{c}}(k) &= \frac{\bar{\mathbf{p}}^{\sf H}\mathbf{U}_{\mathrm{c}}(k)\bar{\mathbf{p}}}{\bar{\mathbf{p}}^{\sf H}\mathbf{V}_{\mathrm{c}}(k)\bar{\mathbf{p}}}, \\ \gamma_k &= \frac{\bar{\mathbf{p}}^{\sf H}\mathbf{U}_{k}\bar{\mathbf{p}}}{\bar{\mathbf{p}}^{\sf H}\mathbf{V}_{k}\bar{\mathbf{p}}},
 \end{align} where
\begin{align}
&{\mathbf{U}}_{\mathrm{c}}(k) =  \, {\mathrm{blkdiag}}\big((\hat{\mathbf{h}}_{k} \hat{\mathbf{h}}_{k}^{\sf H} + {\boldsymbol{\Sigma}}_{k}), \ldots, (\hat{\mathbf{h}}_{k} \hat{\mathbf{h}}_{k}^{\sf H} + {\boldsymbol{\Sigma}}_{k})\big) \nonumber \label{U_c}  \\
& \qquad \qquad+ \frac{\sigma^2}{P}{\mathbf{I}}_{N(K+M+1)}, \\
&{\mathbf{V}}_{\mathrm{c}}(k) =  \, {\mathbf{U}}_{\mathrm{c}}(k) - {\mathrm{blkdiag}}\big({\hat{\mathbf{h}}_{k} \hat{\mathbf{h}}_{k}^{\sf H}}, \mathbf{0},\ldots, {\mathbf{0}}\big),\\
&{\mathbf{U}}_{k} =  \,  {\mathrm{blkdiag}}  \big(  (\mathbf{0},(\hat{\mathbf{h}}_{k} \hat{\mathbf{h}}_{k}^{\sf H} + {\boldsymbol{\Sigma}}_{k}) \ldots, (\hat{\mathbf{h}}_{k} \hat{\mathbf{h}}_{k}^{\sf H} + {\boldsymbol{\Sigma}}_{k})\big)  \nonumber \\
&\qquad \qquad+ \frac{\sigma^2}{P}{\mathbf{I}}_{N(K+M+1)}, \\ 
&{\mathbf{V}}_{k} =  \, {\mathbf{U}}_{k}  - {\mathrm{blkdiag}}  \big(\mathbf{0},  \ldots, \underbrace{\!\hat{\mathbf{h}}_{k} \hat{\mathbf{h}}_{k}^{\sf H}\!}_{(k+1)\text{-th block}}, \ldots, {\mathbf{0}}\big).\label{V_k}
\end{align}
However, we observe that the minimum function in \eqref{P1} is non-smooth, which makes the given problem of the SE maximization problem for RSMA hard to solve. 
To resolve this, we use the LogSumExp technique in \cite{Park:2023, kim:arxiv:24} to approximate the minimum function involved in \eqref{P1}. This gives 
\begin{align}
        \min_{j = 1,...,Q}\{x_j\}  &\approx -\frac{1}{\eta} \log\left(\frac{1}{Q}  \sum_{j = 1}^{Q} \exp\left( {-\eta}{x_j}  \right)\right), \\
        &\triangleq f(\{x_j\}_{j\in[Q]}),
\end{align}
where the larger $\eta$ leads to the tighter approximation. Next, we rewrite the beam gain $G(\mathbf{P};\theta)$ in \eqref{MSE_metric} to express it as a more succinct form using the vector $\bar{\mathbf{p}}$ as follows:
\begin{align}
    G(\mathbf{P};\theta) &= P\cdot\mathbf{a}^{\sf H}(\theta)\mathbf{P}\mathbf{P}^{\sf H}\mathbf{a}(\theta) \\ 
    &= P \cdot \left(\left(\mathbf{a}^{\sf T}(\theta)\mathbf{P}^{*}\right)\otimes\mathbf{a}^{\sf H}(\theta)\right) \mathrm{vec}(\mathbf{P})\label{vec-prop1} \\
    &= P \cdot \left(\left(\left(\mathbf{I}\otimes \mathbf{a}^{\sf T}(\theta)\right)\mathrm{vec}(\mathbf{P}^{*})\right)^{\sf T} \otimes \mathbf{a}^{\sf H}(\theta)\right)\mathrm{vec}(\mathbf{P})\label{vec-prop2} \\
    &= \bar{\mathbf{p}}^{\sf H}\left(P \cdot \mathbf{I}\otimes \mathbf{a}(\theta)\otimes \mathbf{a}^{\sf H}(\theta)\right)\bar{\mathbf{p}} \\
    &\triangleq \bar{\mathbf{p}}^{\sf H}\mathbf{A}(\theta)\bar{\mathbf{p}},\label{beamGain}
\end{align}
where \eqref{vec-prop1}, \eqref{vec-prop2} come from $\mathrm{vec(\mathbf{X}\mathbf{Y}\mathbf{Z})}=(\mathbf{Z}^{\sf T}\otimes \mathbf{X})\mathrm{vec}(\mathbf{Y})$. With \eqref{beamGain}, the beam pattern MSE expression in \eqref{MSE_metric} is rewritten by 
\begin{align}
    \mathrm{MSE}_{\mathrm{r}} = \frac{1}{L}\sum_{u=1}^{L}|\bar{\mathbf{p}}^{\sf H}\mathbf{A}(\theta_{u})\bar{\mathbf{p}}-\bar{\mathbf{p}}^{\sf H}\mathbf{T}(\theta_{u})\bar{\mathbf{p}}|^2, \label{MSE_r}
\end{align}
where $\mathbf{T}(\theta_u) = P \cdot t(\theta_{u})\cdot \mathbf{I}_{N(K+M+1)}$. 

As a result, $\mathscr{P}_1$ is reformulated as $\mathscr{P}_2$:
\begin{align}
    \mathscr{P}_2:& \mathop \textrm{maximize}_{\bar{\mathbf{p}}} \ \bar{R}^{\textrm{LSE}}_{\mathrm{c}}
(\bar{\mathbf{p}}) + \sum_{k=1}^{K}\bar{R}_{k}(\bar{\mathbf{p}})\label{P2}\\
    &\textrm{subject to } \frac{1}{L}\sum_{u=1}^{L}|\bar{\mathbf{p}}^{\sf H}\mathbf{A}(\theta_{u})\bar{\mathbf{p}}-\bar{\mathbf{p}}^{\sf H}\mathbf{T}(\theta_{u})\bar{\mathbf{p}}|^2 \leq T_{\text{mse}},\label{P2-constraint}
\end{align}
where $\bar{R}^{\textrm{LSE}}_{\mathrm{c}} (\bar{\mathbf{p}}), \bar{R}_{k}(\bar{\mathbf{p}})$ are given by
\begin{align}
    \bar{R}^{\textrm{LSE}}_{\mathrm{c}}
(\bar{\mathbf{p}}) &= f\left(\left\{\frac{\bar{\mathbf{p}}^{\sf H}\mathbf{U}_\mathrm{c}(k)\bar{\mathbf{p}}}{\bar{\mathbf{p}}^{\sf H}\mathbf{V}_{\mathrm{c}}(k)\bar{\mathbf{p}}}\right\}_{k\in[K]}\right), \\
\bar{R}_{k}(\bar{\mathbf{p}}) &= \log_2{\frac{\bar{\mathbf{p}}^{\sf H}\mathbf{U}_k\bar{\mathbf{p}}}{\bar{\mathbf{p}}^{\sf H}\mathbf{V}_k\bar{\mathbf{p}}}}.
\end{align}
Note that the power constraint $\|{\mathbf{P}}\|_{\sf F}^2=1$ in $\mathscr{P}_1$ is also removed in $\mathscr{P}_2$, as this constraint will be incorporated into our proposed algorithm.
Notwithstanding this reformulation, directly tackling $\mathscr{P}_2$ is still infeasible because evaluating SINR expressions with \eqref{U_c}-\eqref{V_k} requires the error covariance matrix $\mathbf{\Sigma}_k$, which has not been determined. 
In the next section, we explain challenges in obtaining $\mathbf{\Sigma}_k$ and provide our approach to approximate $\mathbf{\Sigma}_k$.

\section{Error Covariance Matrix Estimation}
Incorporating the error covariance matrix $\mathbf{\Sigma}_k$ in $\mathscr{P}_2$ is important to reflect knowledge of CSI imperfections into the precoder design. 
In conventional approaches, we can obtain $\mathbf{\Sigma}_k$ through (i) Bayesian estimation or (ii) Frequentist approach. 
In Bayesian estimation, we assume that the channel statistics are known in the BS. For example, by modeling the DL channel as a Gaussian distribution with a known covariance, a closed-form expression for the error covariance can be derived, provided that certain CSI feedback is available \cite{FDD-caire-2023, Park:2023}. In the Frequentist approach, we use DL channel error samples to empirically calculate $\mathbf{\Sigma}_k$ as shown in \cite{covApprox}.

In the case considered in this paper, we can easily find that neither approach is feasible. This limitation arises mainly because
the BS cannot obtain CSI error statistics or error samples without feedback \cite{kim:arxiv:24}, whereas we do not rely on any direct CSI feedback. 
To address this, we exploit the CRLB to approximate the CSI error level produced by the UL training process. This approach is validated by the observation that 2D-NOMP can achieve near-optimal MSE \cite{NOMP, Rottenberg:2020}, which implies proximity to the CRLB. 
However, obtaining the CRLB typically requires ground-truth channel parameters \cite{kay1993statistical}, which cannot be obtained in our setup. 
To avoid this problem, we rely on the \textit{ observed Fisher information} \cite{observedFisher}. Specifically, observed Fisher information is the instantaneous amount of information calculated after observing a specific sample, while Fisher information measures the expected amount of information. 
To this end, we define the UL channel parameter set $\mathcal{P}_k$ as 
\begin{align}\mathcal{P}_k &= \{p_{k,1}, \ldots, p_{k,L_k}\} \in \mathbb{C}^{4L_k \times 1}, \\ p_{k, \ell} &= \left\{ \Re\{\alpha^{\mathrm{ul}}_{k, \ell}\}, \Im\{\alpha_{k, \ell}^{\mathrm{ul}}\}, \tau_{k, \ell}^{\mathrm{ul}}, \theta_{k, \ell}^{\mathrm{ul}} \right\}\in \mathbb{C}^{4\times 1}.
\end{align}
Based on this, we find the observed Fisher information by computing the negative Hessian of the log-likelihood composed with an instantaneously estimated parameter $\hat{\mathcal{P}}_k$:
\begin{align}
    \mathbf{I}(\hat{\mathcal{P}}_k) = -\left.\frac{\partial^2\log \mathcal{L}(\mathcal{P}_k; \mathbf{y}_k)}{{\partial \mathcal{P}_k}{\partial \mathcal{P}_k}^{\sf T}}\right|_{{\mathcal{P}}_k=\hat{\mathcal{P}}_k} \in \mathbb{C}^{4L_k\times 4L_k},\label{OFIM}
\end{align}
and the each element of \eqref{OFIM} is given by (See \cite{kim:arxiv:24}),
\begin{align}
\label{Observed Fisher}
&[{\boldsymbol{\mathbf{I}}}(\hat{\mathcal{P}}_k)]_{u, v}=\nonumber \\
&\left.\frac{2}{\sigma^2}\Re\left(\sum_{n, s}\frac{\partial\bar{\mathbf{y}}^{*}_{k,n,s}}{\partial\mathcal{P}_u}\frac{\partial\bar{\mathbf{y}}_{k,n,s}}{\partial\mathcal{P}_v} \right.-\left.(\mathbf{y}_{k,n,s}^{*}-\bar{\mathbf{y}}_{k,n,s}^{*})\frac{\partial^2\bar{\mathbf{y}}_{k,n,s}}{\partial\mathcal{P}_u\partial\mathcal{P}_v}\right)\right|_{{\mathcal{P}}_k=\hat{\mathcal{P}}_k},\end{align}
where $\mathbf{y}_{k, n, s}$ denotes the UL received signal at the $n$-th antenna and $s$-th resource block for user $k$ (in \eqref{y_k original}) and $\bar{\mathbf{y}}_{k, n, s}$ is the UL reconstructed signal defined with the parameter set $\mathcal{P}_k$, i.e., 
\begin{align}
\bar{\mathbf{y}}_{k,n,s} &\triangleq \sum_{\ell=1}^{L_k}\alpha_{k, \ell}^{\mathrm{ul}}[\mathbf{u}(\tau^{\mathrm{ul}}_{k, \ell}, \theta^{\mathrm{ul}}_{k, \ell})]_{n, s},
\end{align}
where $[\mathbf{u}(\tau^{\mathrm{ul}}_{k, \ell}, \theta^{\mathrm{ul}}_{k, \ell})]_{n, s}$ is given in \eqref{u_comp}. 

Interestingly, \cite{efron1978assessing} demonstrated that the inverse of the observed Fisher information provides the optimal performance in estimating the squared error realized. To apply this to the DL case, we utilize the Jacobian transformation as follows:
\begin{align}
    \mathbf{Q}^{\sf H}_k(f)\mathbf{I}^{-1}(\hat{\mathcal{P}}_k)\mathbf{Q}_k(f) = \argmin_{\hat{\mathbf{\Sigma}}_k}\mathbb{E}\left[\sum_{m=1}^N\left[\mathbf{e}_k\mathbf{e}_k^{\sf H}-\hat{\mathbf{\Sigma}}_k\right]_{m,m}^{2}\right],\label{Ofim-Jacobian}
\end{align}
where $\mathbf{Q}^{\sf H}_k(f)$ denotes the Jacobian matrix at the frequency difference $f$ and its derivation is straightforward, thus we omit here (refer to \cite{Rottenberg:2020}). 
To relate this with our case, we can find that the observed Fisher information can be used to minimize the upper bound of error covariance matrix estimation for diagonal elements, i.e.,
\begin{align}\label{OFIM-ECM}
\sum_{m=1}^{N}\left[\mathbb{E}\left[\mathbf{e}_k\mathbf{e}_k^{\sf H}\right] - \hat{\mathbf{\Sigma}}_k \right]_{m,m}^2\leq\mathbb{E}\left[\sum_{m=1}^{N}\left[\mathbf{e}_k\mathbf{e}_k^{\sf H}-\hat{\mathbf{\Sigma}}_k\right]_{m,m}^2\right].
\end{align}
Using this, we estimate the error covariance matrix only with the diagonal component of \eqref{Ofim-Jacobian}.
In~\cite{kim:arxiv:24}, it was shown that the MSE predicted by our method (the trace of \eqref{Ofim-Jacobian}) closely matches the actual MSE. This implies that at least the diagonal components of the error covariance matrix can be accurately estimated with \eqref{Ofim-Jacobian}, allowing $\mathbf{\Sigma}_k$ to be approximated as follows.
\begin{align}
    {\mathbf{\Sigma}}_k &\approx\left(\mathbf{Q}^{\sf H}_k(f)\mathbf{I}^{-1}(\hat{\mathcal{P}}_k)\mathbf{Q}_k(f)\right)\circ\mathbf{I}_N \\
    &=\begin{bmatrix}
\sigma_{k, 1} & 0     & \cdots & 0 \\
0     & \sigma_{k, 2} & \cdots & 0 \\
\vdots& \vdots& \ddots & \vdots \\
0     & 0     & \cdots & \sigma_{k, N}
\end{bmatrix}  = {\mathbf{ \hat \Sigma}_k}, \label{eq:ecm}
\end{align}
where $\sigma_{k, n}$ denotes the MSE at antenna $n$ for user $k$. 
We note that the diagonal error covariance matrix is the case in which uncorrelated channel estimation errors are encountered for each antenna, which is widely assumed for the rich scattering channel where the spatial correlation of the channel is limited \cite{covApprox}. By replacing $\mathbf{\Sigma}_k$ in \eqref{U_c}-\eqref{V_k} with $\hat{\mathbf{\Sigma}}_k$, we are ready to solve problem $\mathscr{P}_2$. 
\textcolor{black}{We note that the robustness of our ECM estimation method can be found in our prior work \cite{kim:arxiv:24}. As shown in Fig. 1 of \cite{kim:arxiv:24}, the estimated ECM closely matches the actual MSE and outperforms the CRLB.}

\section{Precoder Optimization}
In this section, we propose a precoder optimization technique that addresses the problem of maximization of the SE sum under the MSE constraint for the detection of the beam pattern, which guarantees the desired design of the beam pattern. To address this, we first observe that the optimal solutions to problem $\mathscr{P}_2$ adhere to the KKT conditions, which serve as the foundation for our proposed technique. The KKT conditions are summarized as follows:
\begin{itemize}
    \item (i) \textbf{Primal feasibility}: The precoder must satisfy the given constraint, i.e., the inequality constraint in \eqref{P2-constraint}. 
    \item (ii) \textbf{Dual feasibility}:
    For the dual problem, the Lagrange multiplier should be greater than or equal to 0, i.e., $\nu\geq 0$.
    \item (iii) \textbf{Complementary slackness}:
    Relationship between the inequality constraints and their corresponding Lagrange multipliers is given as
    \begin{equation}
    \nu\left(\frac{1}{L}\sum_{\ell=1}^{L}|\bar{\mathbf{p}}^{\sf H}\mathbf{A}(\theta_{\ell})\bar{\mathbf{p}}-\bar{\mathbf{p}}^{\sf H}\mathbf{T}(\theta_{\ell})\bar{\mathbf{p}}|^2 - T_{\text{mse}}\right)=0.
\end{equation}
    \item (iv) \textbf{Stationarity condition}:
    The condition denotes the gradient of the Lagrangian with respect to the optimization variable $\bar{\mathbf{p}}$ is equal to zero. 
\end{itemize}
Upon this, we summarize the stationarity condition for the dual problem of $\mathscr{P}_2$ with the following lemma.
\begin{lemma}
    The stationarity condition for the dual problem of $\mathscr{P}_2$ holds if the following equation is satisfied:
    \begin{align}
        \mathbf{L}(\bar{\mathbf{p}}, \nu)\bar{\mathbf{p}} = \zeta(\bar{\mathbf{p}}, \nu)\mathbf{R}(\bar{\mathbf{p}}, \nu)\bar{\mathbf{p}},\label{FONC}
    \end{align}
    where
    \begin{align}
    &\mathbf{L}(\bar{\mathbf{p}}, \nu) =\zeta_{\mathrm{num}}(\bar{\mathbf{p}}, \nu)\times\nonumber\\ 
    &\Bigg[\sum_{k=1}^{K}\left\{\frac{\exp\left(-\eta\frac{\bar{\mathbf{p}}^{\sf{H}}\mathbf{U}_{\mathrm{c}}(k)\bar{\mathbf{p}}}{\bar{\mathbf{p}}^{\sf{H}}\mathbf{V}_\mathrm{c}(k)\bar{\mathbf{p}}}\right)}{\sum_{j=1}^{K}\exp\left(-\eta\log_2\frac{\bar{\mathbf{p}}^{\sf{H}}\mathbf{U}_{\mathrm{c}}(j)\bar{\mathbf{p}}}{\bar{\mathbf{p}}^{\sf{H}}\mathbf{V}_{\mathrm{c}}(j)\bar{\mathbf{p}}}\right)}\right\}\frac{\mathbf{V}_{\mathrm{c}}(k)}{\bar{\mathbf{p}}^{\sf H}\mathbf{V}_{\mathrm{c}}(k)\bar{\mathbf{p}}}+\sum_{k=1}^K\left(\frac{\mathbf{V}_k}{\bar{\mathbf{p}}^{\sf H}\mathbf{V}_{k}\bar{\mathbf{p}}}\right) \label{L(p)}\nonumber\\
    &\quad +\frac{4\nu\log2}{LP^2}\sum^L_{u=1}\left((\bar{\mathbf{p}}^{\sf H}\mathbf{A}(\mathrm{\theta}_u)\bar{\mathbf{p}})\mathbf{A}(\mathrm{\theta}_u) +(\bar{\mathbf{p}}^{\sf H}\mathbf{T}(\mathrm{\theta}_u)\bar{\mathbf{p}})\mathbf{T}(\mathrm{\theta}_u)\right)\bar{\mathbf{p}}\Bigg],\\
    &\mathbf{R}(\bar{\mathbf{p}}, \nu) = \zeta_{\mathrm{den}}(\bar{\mathbf{p}}, \nu) \times\nonumber\\ 
    &\Bigg[\sum_{k=1}^{K}\left\{\frac{\exp\left(-\eta\frac{\bar{\mathbf{p}}^{\sf{H}}\mathbf{U}_{\mathrm{c}}(k)\bar{\mathbf{p}}}{\bar{\mathbf{p}}^{\sf{H}}\mathbf{V}_\mathrm{c}(k)\bar{\mathbf{p}}}\right)}{\sum_{j=1}^{K}\exp\left(-\eta\log_2\frac{\bar{\mathbf{p}}^{\sf{H}}\mathbf{U}_{\mathrm{c}}(j)\bar{\mathbf{p}}}{\bar{\mathbf{p}}^{\sf{H}}\mathbf{V}_{\mathrm{c}}(j)\bar{\mathbf{p}}}\right)}\right\}\frac{\mathbf{U}_{\mathrm{c}}(k)}{\bar{\mathbf{p}}^{\sf H}\mathbf{U}_{\mathrm{c}}(k)\bar{\mathbf{p}}}\nonumber+\sum_{k=1}^K\left(\frac{\mathbf{U}_k}{\bar{\mathbf{p}}^{\sf H}\mathbf{U}_{k}\bar{\mathbf{p}}}\right) \nonumber\\
    &\quad +\frac{4\nu\log2}{LP^2}\sum^L_{u=1}\left((\bar{\mathbf{p}}^{\sf H}\mathbf{A}(\mathrm{\theta}_u)\bar{\mathbf{p}})\mathbf{T}(\mathrm{\theta}_u) +(\bar{\mathbf{p}}^{\sf H}\mathbf{T}(\mathrm{\theta}_u)\bar{\mathbf{p}})\mathbf{A}(\mathrm{\theta}_u)\right)\Bigg],\label{R(p)}
    \end{align}
    in which $\nu$ in \eqref{L(p)} and \eqref{R(p)} denotes the Lagrangian multiplier, and we define $\zeta(\bar{\mathbf{p}}, \nu)$ as
    \begin{align}
        \zeta(\bar{\mathbf{p}}, \nu)&=\left \{\frac{1}{K}\sum _{k = 1}^{K} {\exp \!\left ({-\eta\log _{2} \left ({\frac {\bar {\mathbf{p}}^{\sf H} \mathbf{U}_{\mathrm{c}}(k)\bar {\mathbf{p}}}{\bar {\mathbf{p}}^{\sf H} \mathbf{V}_{\mathrm{c}}(k)\bar{\mathbf{p}} }}\right) }\right)}\right \}^{-\frac{1} {\eta\log _{2} e}}\nonumber\\ 
        &\quad\times\prod_{k=1}^K\frac{\bar{\mathbf{p}}^{\sf H}\mathbf{U}_k\bar{\mathbf{p}}}{\bar{\mathbf{p}}^{\sf H}\mathbf{V}_k\bar{\mathbf{p}}} \times 2^{\nu\left(\frac{1}{LP^2}\sum_{u=1}^{L}\left|\bar{\mathbf{p}}^{\sf{H}}\mathbf{A}(\theta_u)\bar{\mathbf{p}} - \bar{\mathbf{p}}^{\sf{H}}\mathbf{T}(\theta_u)\bar{\mathbf{p}}\right|^2-\frac{T_{\mathrm{mse}}}{P^2}\right)}\label{zeta}\\
        &=\frac{\zeta_{\mathrm{num}}(\bar{\mathbf{p}}, \nu)}{\zeta_{\mathrm{den}}(\bar{\mathbf{p}}, \nu)}.
    \end{align}
\end{lemma}

\begin{algorithm}
\caption{GPI-ISAC-RS Algorithm}
\SetAlgoLined
\KwIn{UL received signal, $\mathbf{y}_k\in \mathbb{C}^{MN\times 1}$}
\KwOut{DL precoder, $\bar{\mathbf{p}}^\ast \in \mathbb{C}^{N(K+M+1)\times 1}$}

\textbf{Initialize:} $\bar{\mathbf{p}}^{(0)}$, $\nu^{(0)}$\;
Set each iteration count as $t = 1, n = 1$\;


Reconstruct the DL channel $\hat{\mathbf{h}}_k, \forall k$ via Section III\;
Estimate the error covariance $\hat{\mathbf{\Sigma}}_k, \forall k$ via Section IV\;

\While{$|\nu^{(n)} - \nu^{(n-1)}| \geq \epsilon_{\nu}$ \textbf{or} $n \leq n_{\text{max}}$}{
    Set the iteration count $t = 1$\;
    \While{$\|\bar{\mathbf{p}}^{(t)} - \bar{\mathbf{p}}^{(t-1)}\| \geq \epsilon_{p}$ \textbf{or} $t \leq t_{\text{max}}$}{
    Construct matrix $\mathbf{L}(\bar{\mathbf{p}}^{(t-1)},\nu^{(n)})$ with (33)\;
    Construct matrix $\mathbf{R}(\bar{\mathbf{p}}^{(t-1)}, \nu^{(n)})$ with (34)\;
    Compute matrix $\mathbf{K}(\bar{\mathbf{p}}^{(t-1)}, \nu^{(n)}) \leftarrow \mathbf{R}^{-1}(\bar{\mathbf{p}}^{(t-1)}, \nu^{(n)})\mathbf{L}(\bar{\mathbf{p}}^{(t-1)}, \nu^{(n)})$\;
    Update $\bar{\mathbf{p}}(t)$ as:
    \[
    \bar{\mathbf{p}}^{(t)} \leftarrow \frac{\mathbf{K}(\bar{\mathbf{p}}^{(t-1)}, \nu^{(n)})\bar{\mathbf{p}}^{(t-1)}}{\|\mathbf{K}(\bar{\mathbf{p}}^{(t-1)}, \nu^{(n)})\bar{\mathbf{p}}^{(t-1)}\|},
    \]
    $t \leftarrow t + 1$\;
}
    \eIf{$\zeta_{\text{mse}} > \frac{1}{L} \sum_{\ell=1}^{L} | G(\mathbf{P}; \theta_{\ell})- P\cdot t(\theta_{\ell})|^2$}{
        $\nu^{(n+1)} \leftarrow \frac{\nu^{(n)} + \nu_{\text{min}}}{2}$\;
    }{
        $\nu^{(n+1)} \leftarrow \frac{\nu^{(n)} + \nu_{\text{max}}}{2}$\;
    }
    $n \leftarrow n + 1$\;
}

$\bar{\mathbf{p}}^\ast \leftarrow \bar{\mathbf{p}}(t-1)$\;
\Return $\bar{\mathbf{p}}^\ast$\;

\end{algorithm}

However, note that even if we can find a precoding vector that satisfies the KKT conditions in (i)-(iv), it still does not imply that it is the global optimal point, since the underlying problem of \eqref{P2} remains non-convex.
Instead, only a necessary condition of the local optimal point is ensured by jointly finding the precoding vector $\bar{\mathbf{p}}$ and the Lagrangian multipler $\nu$ that satisfies the KKT conditions. 
Furthermore, it should be noted that the Lagrange multiplier $\nu$ affects the optimal precoder $\bar{\mathbf{p}}^{\ast}$, and vice versa, thus the interdependence makes it challenging to optimize $\bar{\mathbf{p}}$ and $\nu$ simultaneously.

To address this problem, we first highlight the relationship between the objective function in \eqref{P2} and $\zeta(\bar{\mathbf{p}}, \nu)$ in \eqref{FONC}, under the assumption that the KKT conditions hold. Specifically, if a precoder and Lagrange multiplier pair $(\bar{\mathbf{p}}^{\ast}, \nu^{\ast})$ satisfies the complementary slackness, then the following holds: \begin{align} \log_2\zeta(\bar{\mathbf{p}}^{\ast}, \nu^{\ast}) = \bar{R}^{\textrm{LSE}}_{\mathrm{c}} (\bar{\mathbf{p}}^{\ast}) + \sum_{k=1}^K\bar{R}_{k}(\bar{\mathbf{p}}^{\ast}), \end{align} where the MSE constraint term in \eqref{zeta} vanishes, and $\zeta(\bar{\mathbf{p}}^{\ast}, \nu^{\ast})$ is simply the sum SE, i.e., our objective function. Therefore, insofar as the KKT conditions are satisfied, we just need to find $(\bar{\mathbf{p}}^{\ast}, \nu^{\ast})$ that maximizes $\zeta(\bar{\mathbf{p}}^{\ast}, \nu^{\ast})$, which is equivalent to maximizing the objective of $\mathscr{P}_2$. 

Based on this, we revisit the stationarity condition in Lemma 1 and interpret the problem of maximizing $\zeta(\bar{\mathbf{p}}, \nu)$ as a form of NEPv \cite{Park:2023, choi:twc:24}. We focus on $\zeta(\bar{\mathbf{p}}, \nu)$ by multiplying the inverse of $\mathbf{R}(\bar{\mathbf{p}}, \nu)$ on both sides of \eqref{FONC}, which is given by 
\begin{align}
    \mathbf{R}^{-1}(\bar{\mathbf{p}}, \nu)\mathbf{L}(\bar{\mathbf{p}}, \nu)\bar{\mathbf{p}} = \zeta(\bar{\mathbf{p}}, \nu)\bar{\mathbf{p}}.\label{eq:stationarity}
\end{align}
With this, assuming only the update of $\bar{\mathbf{p}}$ while keeping $\nu$ fixed, we summarize our precoder design objective as follows:

\begin{proposition}
  Suppose a Lagrange multiplier $\nu^{\dagger} \geq 0$ is given, and define the optimal precoder $\mathbf{p}^{\dagger}$ as the vector that maximizes the objective function $\zeta(\bar{\mathbf{p}}, \nu^{\dagger})$ for the fixed $\nu^{\dagger}$, under the stationarity condition in \eqref{eq:stationarity}, which satisfies
\[
\mathbf{R}^{-1}(\bar{\mathbf{p}}, \nu^{\dagger})\mathbf{L}(\bar{\mathbf{p}}, \nu^{\dagger})\bar{\mathbf{p}} = \zeta(\bar{\mathbf{p}}, \nu^{\dagger})\bar{\mathbf{p}}.
\]  
Then, $\mathbf{p}^{\dagger}$ is the leading eigenvector of the matrix $\mathbf{K}(\bar{\mathbf{p}}, \nu^{\dagger}) \triangleq \mathbf{R}^{-1}(\bar{\mathbf{p}}, \nu^{\dagger})\mathbf{L}(\bar{\mathbf{p}}, \nu^{\dagger})$.

\end{proposition}
This proposition provides an useful insight, as it allows the problem of maximizing \(\zeta(\bar{\mathbf{p}}, \nu^{\dagger})\) to be interpreted as a form of NEPv. Specifically, \(\zeta(\bar{\mathbf{p}}, \nu^{\dagger})\) is cast as a nonlinear eigenvalue of the matrix \(\mathbf{K}(\bar{\mathbf{p}}, \nu^{\dagger})\) and $\bar{\mathbf{p}}$ is the corresponding eigenvector. 


Inspired by \cite{Park:2023, choi:twc:24}, we can find the leading eigenvector of $\mathbf{K}(\bar{\mathbf{p}}, \nu^{\dagger})$ with the generalized power iteration (GPI) method. In particular, we repeatedly update the precoder to be projected into the maxtrix $\mathbf{K}(\bar{\mathbf{p}}, \nu^{\dagger})$ and be normalized to ensure the original power constraint in \eqref{P0_TX_Power}. Specifically, at $t$-th iteration, we update the precoder as
\begin{align}
    \bar{\mathbf{p}}^{(t)} \leftarrow \frac{\mathbf{K}(\bar{\mathbf{p}}^{(t-1)}, \nu^{\dagger})\bar{\mathbf{p}}^{(t-1)}}{\|\mathbf{K}(\bar{\mathbf{p}}^{(t-1)}, \nu^{\dagger})\bar{\mathbf{p}}^{(t-1)}\|}, \label{eq:GPI}
\end{align}
and this process is repeated until the termination criterion is met. For example, we expect that it is converged when $|\bar{\mathbf{p}}^{(t)}-\bar{\mathbf{p}}^{(t-1)}|<\epsilon, $ where $\epsilon$ is small enough. 

To understand how the proposed method maximizes the nonlinear eigenvalue $\zeta(\bar{\mathbf{p}}, \nu^{\dagger})$, we represent the inner product of denominator in \eqref{eq:GPI} with an arbitrary precoder $\bar{\mathbf{p}}$ as follows:
\begin{align}
    &(\mathbf{K}(\bar{\mathbf{p}}^{(t-1)}, \nu^{\dagger})\bar{\mathbf{p}}^{(t-1)})^{\sf H}\bar{\mathbf{p}}\notag = (\mathbf{K}(\bar{\mathbf{p}}^{(t-1)}, \nu^{\dagger})\bar{\mathbf{p}}^{\dagger})^{\sf H}\bar{\mathbf{p}}\\
    &+(\bar{\mathbf{p}}^{(t-1)}-\bar{\mathbf{p}}^{\dagger})^{\sf H}\nabla_{\bar{\mathbf{p}}^{(t-1)}}(\mathbf{K}(\bar{\mathbf{p}}^{\dagger}, \nu^{\dagger})\bar{\mathbf{p}}^{\dagger})\bar{\mathbf{p}}
    +o(\|\bar{\mathbf{p}}^{(t-1)}-\bar{\mathbf{p}}^{\dagger}\|), \label{eq:Taylor}
\end{align}
Assuming that there exists a set of basis $\{\bar{\mathbf{p}}^{\dagger}, \mathbf{v}_2, \cdots,\mathbf{v}_{N(K+M+1)}\}$ for the matrix $\mathbf{K}(\bar{\mathbf{p}}^{(t-1)}, \nu^{\dagger})$, we also rewrite \eqref{eq:Taylor} as
\begin{align}
    &(\mathbf{K}(\bar{\mathbf{p}}^{(t-1)}, \nu^{\dagger})\bar{\mathbf{p}}^{(t-1)})^{\sf H}\bar{\mathbf{p}}= \alpha_1(\mathbf{K}(\bar{\mathbf{p}}^{(t-1)}, \nu^{\dagger})\bar{\mathbf{p}}^{(t-1)})^{\sf H}\bar{\mathbf{p}}^{\dagger}\notag\\ &+ \sum_{i=2}^{N(K+M+1)}\alpha_i(\mathbf{K}(\bar{\mathbf{p}}^{(t)}, \nu^{\dagger})\bar{\mathbf{p}}^{(t-1)})^{\sf H})\mathbf{v}_{i},
\end{align}
since $\bar{\mathbf{p}}$ can be expressed as $\alpha_1\bar{\mathbf{p}}^{\dagger} + \sum_{i=2}^{N(K+M+1)}\alpha_i\mathbf{v}_i$ by using a basis set. 

Now, considering that $\mathbf{K}(\bar{\mathbf{p}}^{\dagger}, \nu^{\dagger})\bar{\mathbf{p}}^{\dagger} = \lambda^{\dagger}\bar{\mathbf{p}}^{\dagger}$ holds, we obtain from \eqref{eq:Taylor}
\begin{align}
    [(\mathbf{K}(\bar{\mathbf{p}}^{(t-1)}, \nu^{\dagger})\bar{\mathbf{p}}^{(t-1)})^{\sf H}\bar{\mathbf{p}}^{\dagger}]^2 = [\lambda^{\dagger}+o(\|\bar{\mathbf{p}}^{(t-1)}-\bar{\mathbf{p}}^{\dagger}\|)]^2.\label{eq:leading}
\end{align}
We also derive an upper bound for the sum of the inner product of the remaining orthonormal basis, $\{\mathbf{v}_i\}_{i=2,\ldots,N(K+M+1)}$, as follows. 
\begin{align}
    &\sum_{i=2}^{N(K+M+1)}[(\mathbf{K}({\bar{\mathbf{p}}}^{(t-1)}, \nu^{\dagger})\bar{\mathbf{p}}^{(t-1)})^{\sf H}\mathbf{v}_i]^2\\ &\leq \sum_{i=2}^{N(K+M+1)}[\lambda_i^2(({\bar{\mathbf{p}}}^{(t-1)})^{\sf H}\mathbf{v}_i)^2+2\lambda_i((\bar{\mathbf{p}}^{(t-1)})^{\sf H})\mathbf{v}_i)\notag\\
    &\qquad \qquad \qquad \qquad \qquad \qquad +o(\|\bar{\mathbf{p}}^{(t-1)}-\bar{\mathbf{p}}^{\dagger}\|)^2]\\
    &\leq [\lambda_2\|{\bar{\mathbf{p}}}^{(t-1)}-{\bar{\mathbf{p}}}^{\dagger}\|+o(\|\bar{\mathbf{p}}^{(t-1)}-\bar{\mathbf{p}}^{\dagger}\|)]^2,\label{eq:non-leading}
\end{align}
which is based on the following fact: 
\begin{align}
    &\sum_{i=2}^{N(K+M+1)}(({\bar{\mathbf{p}}}^{(t-1)})^{\sf H}\mathbf{v}_i)^2 = 1 - (({\bar{\mathbf{p}}}^{(t-1)})^{\sf H}{\bar{\mathbf{p}}}^{\dagger})^2 \\
    &\leq 2(1-({\bar{\mathbf{p}}}^{(t-1)})^{\sf H}{\bar{\mathbf{p}}}^{\dagger}) \leq \|{\bar{\mathbf{p}}}^{(t-1)}-{\bar{\mathbf{p}}}^{\dagger}\|^2.
\end{align}
By comparing \eqref{eq:leading} with \eqref{eq:non-leading}, 
given that ${\bar{\mathbf{p}}}^{(t-1)}$ is close enough to ${\bar{\mathbf{p}}}^{\dagger}$, 
iteratively projecting the precoder ${\bar{\mathbf{p}}}^{(t-1)}$ onto $\mathbf{K}({\bar{\mathbf{p}}}^{(t-1)}, \nu^{\dagger})$ makes the non-leading eigenvectors $\{\mathbf{v}_i\}_{i=2,\ldots,N(K+M+1)}$ vanish, while the leading eigenvector ${\bar{\mathbf{p}}}^{\dagger}$ remains, as $|\lambda^{\dagger}| > |\lambda_i|,\; \forall i \geq 2$. 
Therefore, our proposed algorithm approaches the leading eigenvector $\bar{\mathbf{p}}^{\dagger}$.

Recall that we have assumed a fixed Lagrange multiplier $\nu = \nu^{\dagger}$, which is valid as long as the KKT condition holds. To ensure this, we now explore how to update $\nu$. First, $\nu$ must be non-negative due to dual feasibility, and it must also satisfy complementary slackness when multiplied by the inequality constraints in the dual problem. Meanwhile, the gradient component related to the MSE constraint in \eqref{L(p)}, \eqref{R(p)} is linearly proportional to $\nu$, which implies that the design objectives can be adjusted based on the value of $\nu$. We summarize this in the following remark.

\begin{remark}(Trade-off between communication and sensing)
\normalfont Notably, Lagrange multiplier $\nu$ dictates the design direction between communication and sensing. If we find a precoder $\bar{\mathbf{p}}$ that maximizes $\zeta(\bar{\mathbf{p}}, \nu)$ with $\nu$ close to zero, it indicates that the given MSE constraint is loose, allowing us to focus on optimizing the precoder to maximize the sum SE without considering the sensing constraint. In contrast, when we increase the value of $\nu$, we obtain the precoder solution with more highlights in the MSE constraint for the dual problem, which is indeed not favorable to communication performance due to the trade-off relationship between sensing and communication. 
\end{remark}

Following the intuition of the above remark, we choose $\nu$ to be as small as possible, provided that it satisfies the MSE constraints of the detection beam pattern. In this regard, we employ a binary tree search method. Specifically, we begin with a $\nu$ value of zero and derive the optimal precoder that maximizes the sum SE using the NEPv interpretation. We then verify whether this solution meets the MSE constraint. If the feasibility conditions are not satisfied, we increase the $\nu$ value to more account for the MSE constraint in the next iteration. In contrast, if the solution meets the constraints, we attempt to decrease the $\nu$ value to achieve smaller feasible values. Through this approach, we effectively find the smallest $\nu$ that satisfies all KKT conditions, which we denote $\nu^{\ast}$. With this, we repeat the process in \eqref{eq:GPI} and obtain the corresponding precoding vector $\bar{\mathbf{p}}^{\ast}$. 
In conclusion, we alternatively optimize $(\bar{\mathbf{p}}, \nu)$ with our proposed method, which is summarized in \textbf{Algorithm 1}.

\textcolor{black}{\begin{remark}(Comparison to conventional codebook-based CSI acquisition)
\normalfont 
Following 5G NR, acquiring DL CSI in FDD systems typically involves the following steps: (i) the BS transmits CSI-RS; (ii) the user estimates the DL CSI and computes a CSI report based on a predefined codebook; (iii) the user sends this report back to the BS over the UL control channel; and (iv) the BS decodes the feedback to recover the DL CSI. This process introduces considerable latency and signaling overhead (approximately 9 ms in the worst case), which poses challenges for time-sensitive ISAC applications (e.g., autonomous driving). 
In contrast, our framework reconstructs DL CSI directly from UL reference signals, eliminating the need for DL training, user-side computation, and UL feedback. 
Nonetheless, we emphasize that our framework is not intended to completely replace existing codebook-based methods. Since it does not rely on direct CSI feedback, our approach inevitably involves CSI estimation errors, whereas codebook-based methods are generally more suitable for acquiring accurate DL CSI. 
However, by avoiding the significant overhead and latency, it provides a compelling alternative for scenarios demanding ultra-low latency and reduced signaling FDD ISAC applications such as autonomous driving and real-time sensing. 
\end{remark}}

\textcolor{black}{\begin{remark}(Key assumptions and potential limitations)
\normalfont 
We summarize key assumptions required for our method to operate efficiently. 
First, our method relies on the assumption of partial reciprocity between UL and DL channels. Specifically, reciprocal  propagation delay, angular parameter, channel path number, and correlated DL-UL channel gains are necessary to reconstruct the DL CSI from UL reference signals. As explained in Remark \ref{remark_partial}, these hold under moderate frequency separations but may be violated under large separations \cite{shakya2024urban}. 
Accordingly, the performance of our method degrades as the frequency separation between UL and DL increases.
In addition, as 2D-NOMP is exploited, a sparse multipath environment is needed where the number of path is relatively smaller than the number of antennas and subcarriers. 
Further, sufficient computational resources are necessary to perform CSI reconstruction within the channel coherence time, along with sufficiently high UL SNR for reliable DL CSI reconstruction. 
\end{remark}}

\section{Simulation Results}
\begin{figure}[t]
\centering

\begin{subfigure}
    \centering
    \includegraphics[width=\linewidth]{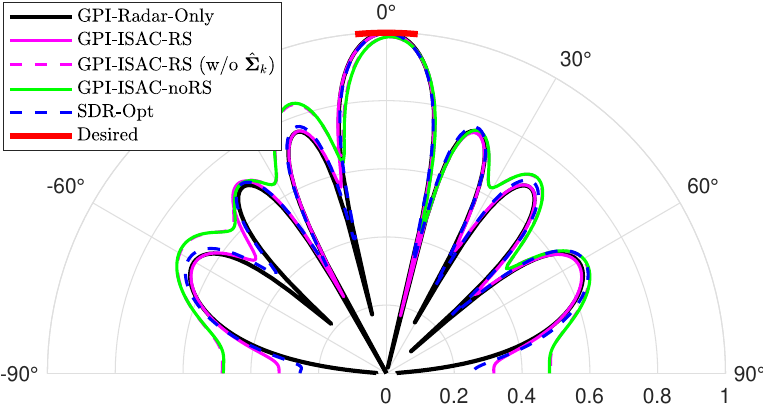}
    \caption*{(a) Single-target case, target is at 0\degree}
    \label{fig:h1}
\end{subfigure}

\vspace{1em}

\begin{subfigure}
    \centering
    \includegraphics[width=\linewidth]{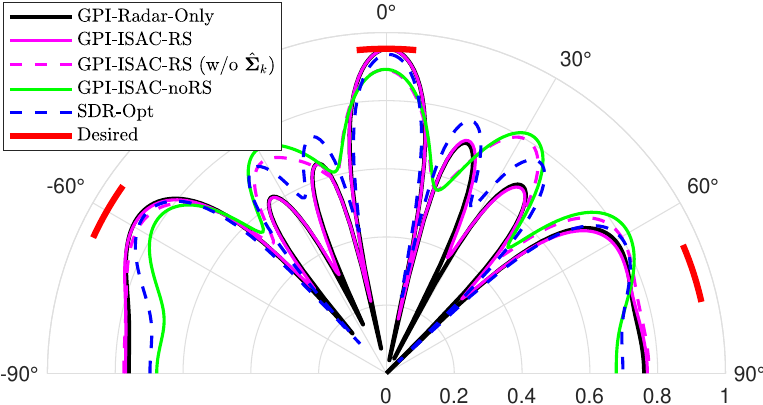}
    \caption*{(b) Multi-target case, targets are at -60\degree, 0\degree, 72\degree}
    \label{fig:h2}
\end{subfigure}

\vspace{1em}

\color{black}
\caption*{\textcolor{black}{Sidelobe suppression (in dB scale)}}
\begin{center}
\begin{tabular}{ccc}
\hline
\textbf{Method} & \textbf{Single-target} & \textbf{Multi-target} \\
\hline
GPI-Radar-Only & -10.8 & -10.56 \\
GPI-ISAC-RS & -10.2 & -10.14 \\
GPI-ISAC-RS (w/o $\hat{\mathbf{\Sigma}}_k$) & -9.92 & -3.06 \\
GPI-ISAC-noRS & -7.61 & -2.58 \\
SDR-Opt & -9.73 & -4.08 \\
\hline
\end{tabular}
\end{center}

\caption{Normalized beam patterns and corresponding sidelobe suppression evaluations for $N=8, K=4, M=4$.}
\label{fig:beam_and_table}
\end{figure}

In this section, we demonstrate the performance of the proposed method in various settings. First, for comparison, we present existing state-of-the-art ISAC precoding methods as follows: 
\begin{itemize}
    \item \textbf{GPI-ISAC (no RS):} As a recently proposed ISAC framework under imperfect CSI \cite{choi:twc:24}, it considers the problem in \eqref{P1}; yet with conventional SDMA. 
    \item \textbf{Semi-definite relaxation (SDR)-based method:} As shown in \cite{liu:tsp:18, liu:tsp:20}, this method is based on the SDR technique, which solves the following problem:
    \begin{align}
        &\mathop\textrm{minimize}_{{\mathbf{P}}} \mathrm{MSE}_{\mathrm{r}} \label{SDR-obj} \\
        &\textrm{subject to }  \mathrm{SINR}_k \geq \Gamma_k, \|\mathbf{P}\|^2_F = 1, \label{SDR-const}
    \end{align}
thereafter we perform a rank-1 approximation to obtain feasible precoding vectors from the positive semi-definite matrix. 
In \eqref{SDR-obj}, it can be observed that the constraint and objective function are changed compared to the case of \(\mathscr{P}_1\) in \eqref{P1}, indicating that the SDR-based method is a radar-centric approach as explained in \cite{choi:twc:24}. In \eqref{SDR-const}, we determine the SINR constraint \(\Gamma_k\) from GPI-ISAC for a fair comparison. If the algorithm fails to find a precoder set that satisfies the constraint, we reduce \(\Gamma_k\) for all \(k\) by 5\% to make the problem feasible.
\end{itemize}
\textcolor{black}{To evaluate the effectiveness of the proposed approach under realistic conditions, we conduct simulations assuming the UL and DL carrier frequencies are set to $f_{\mathrm{c}}^{\mathrm{ul}} = 7.25$ GHz and $f_{\mathrm{c}}^{\mathrm{dl}} = 7.75$ GHz, respectively.}

\subsection{Sensing meam shaping for desired target}

Fig.~2 illustrates the beam patterns used to evaluate sensing performance for each technique (with $\nu = \nu_{\textrm{max}}$ in GPI-based methods). We consider both single- and multi-target scenarios, with target directions indicated in the figure. The Radar-only approach introduced in \cite{choi:twc:24} shows the lowest sidelobe levels and the highest beam gain at the target angle in both cases, as it fully prioritizes sensing without accounting for communication. In contrast, ISAC-based methods generally exhibit higher sidelobe levels. Among them, the proposed GPI-ISAC-RS achieves the lowest sidelobe levels while maintaining strong main-lobe gain, indicating superior sensing performance. When the ECM is not applied in our method, both sidelobe levels and beam gain degrade, leading to reduced sensing performance. This demonstrates that ECM estimation, though originally intended to improve communication metrics, also benefits sensing performance in ISAC systems by enabling more efficient resource sharing. Furthermore, the use of RS contributes to additional gains, as the common stream in \eqref{MSE_r} reinforces the sensing objective in the sensing-centric regime.

\textcolor{black}{
To complement the beam pattern analysis with quantitative metrics, the table below the figures presents the sidelobe suppression values for each method shown in Fig.~2. These values represent the dB difference between the main-lobe peak and the highest sidelobe, based on the original beam patterns already normalized in the dB scale. As the table shows, our ECM-aided GPI-ISAC-RS achieves the greatest suppression across both scenarios, aligning well with the visual trends in Fig.~2.
}
\begin{figure} \centering
  \begin{subfigure}
  \centering{
  \includegraphics[width=0.9\columnwidth]
  {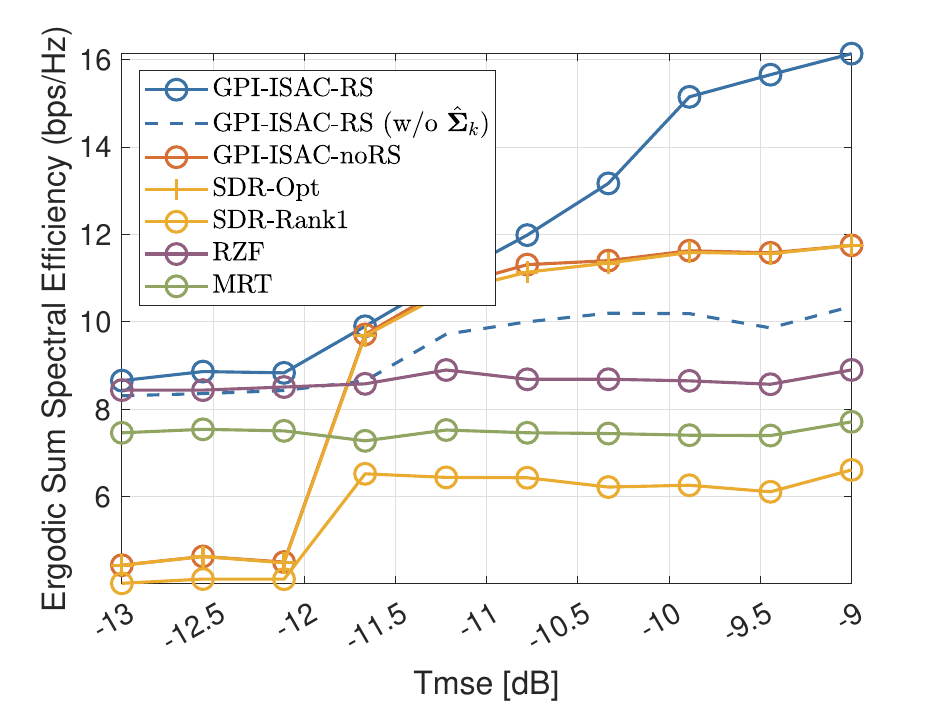} 
 \caption*{(a) Ergodic sum SE}}
  \end{subfigure}
 \begin{subfigure}
 \centering{
 \includegraphics[width=0.9\columnwidth]
 {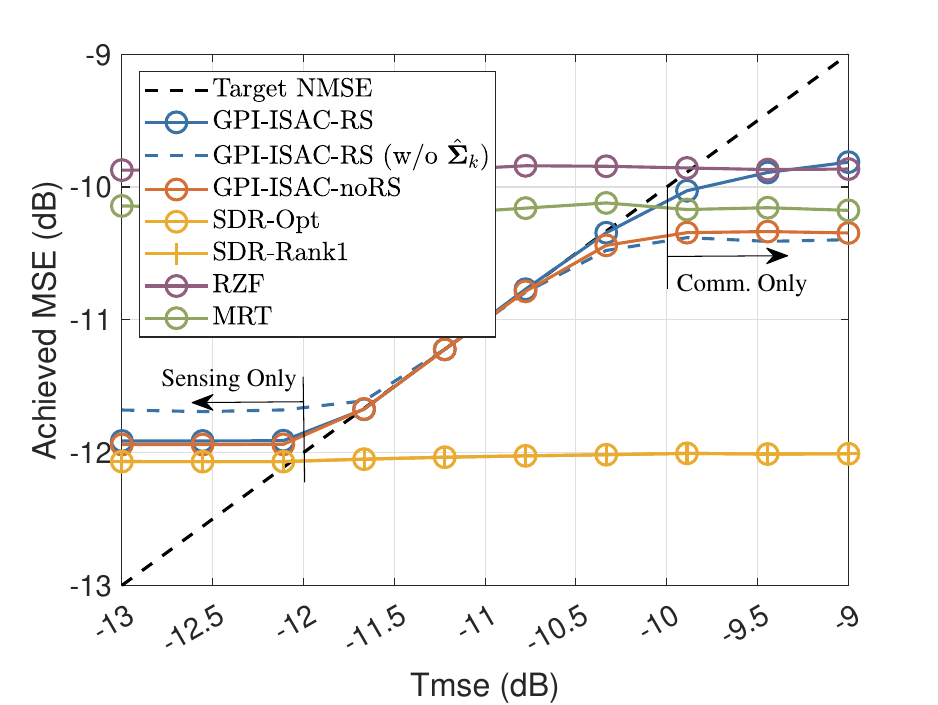}
  \caption*{(b) Achieved MSE}}
  \end{subfigure}
\caption{(a) Achieved MSE and (b) Ergodic sum SE with $N = 8, K = 4, M = 4, \eta_{k, \ell} = 0.9 \ \forall (k, \ell)$, the SNR (= $P/\sigma^2$) of 35 dB is assumed.}\label{figure_targetMSE_4}
\end{figure}
\subsection{Performance over target MSE}
\subsubsection{SE vs target MSE}
In Fig. \ref{figure_targetMSE_4}, we examine the ergodic sum SE and achieved MSE in \eqref{MSE_metric} over the target MSE, $T_{\mathrm{mse}}$. For a comprehensive comparison, we include communication-only methods such as MRT and RZF alongside ISAC-supported benchmarks. 
As shown in Fig. \ref{figure_targetMSE_4}-(a), a lower MSE constraint results in a reduced SE for all ISAC methods, highlighting the trade-off relationship between sensing and communication. That is, focusing on beam pattern matching for a specific target with higher accuracy reduces communication performance, while prioritizing communication performance compromises sensing accuracy. Despite this, the proposed GPI-ISAC-RS achieves the best SE performance across all given target MSE levels. This relative performance gain becomes particularly pronounced at lower target MSE levels, which correspond to the sensing-oriented operational region. This indicates that the proposed RS-based method can maintain communication performance even when spatial resources are allocated to sensing targets \cite{xu:jstsp:22}. In particular, we also observe that the proposed error covariance estimation method is crucial for achieving the RS gain. Without it, performance decreases significantly, even lower than that of the GPI-ISAC case in most of the target MSE region.

\subsubsection{Achieved MSE vs target MSE}
The foundation of the SE gain becomes evident when analyzing the trends in the actual MSE achieved for each target MSE. In particular, when the radar beam pattern constraint is relaxed (higher $T_{\mathrm{mse}}$), as shown in Fig. 3-(b), the actual MSE achieved by the proposed method adapts more effectively to the target MSE, contributing to SE gains. In other words, compared to the no-RS method, the additional common message stream in the proposed technique is used to maximize communication performance, resulting in a higher actual MSE but ultimately improves SE gains. 
On the other hand, when the constraint of the beam pattern is difficult to satisfy (lower $T_{\mathrm{mse}}$), the proposed method does not show a significant improvement in the MSE performance compared to the case without RS. However, this radar-centric precoding approach can cause interference with communication functionality, which RS effectively manages to achieve SE gains. Crucially, it is observed that failing to estimate error covariance leads to suboptimal MSE performance. This highlights that accurately estimating the error covariance is of importance not only for communication performance but also for sensing performance.

\begin{figure} \centering
 \begin{subfigure}
 \centering{
 \includegraphics[width=0.9\columnwidth]{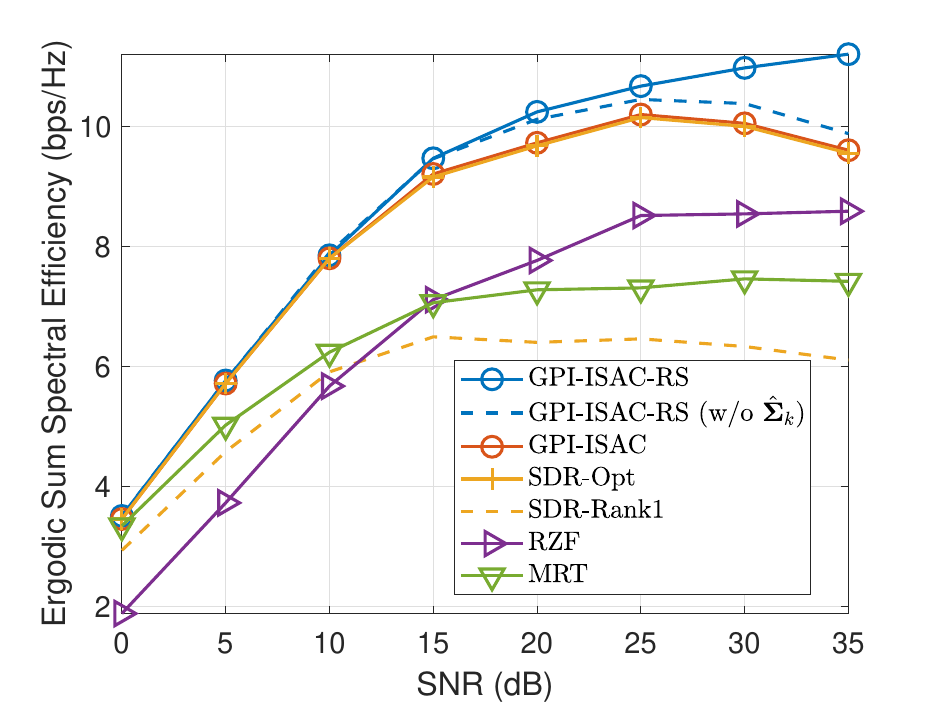} 
 \caption*{(a) Ergodic sum SE}}
 \label{fig:h1}
  \end{subfigure}
  \begin{subfigure}
  \centering{
  \includegraphics[width=0.9\columnwidth]{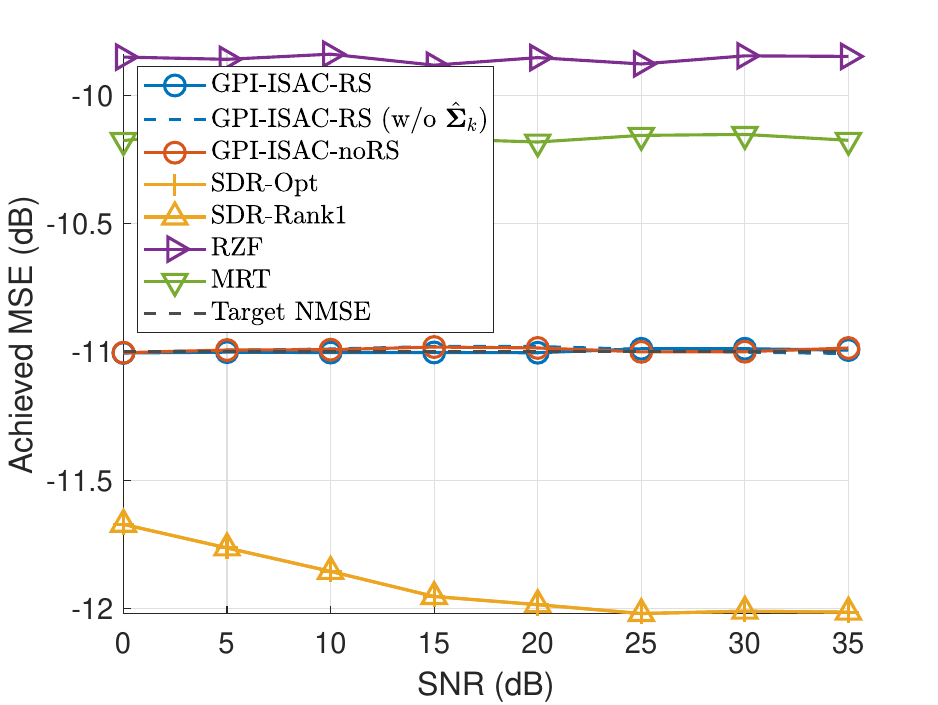}
  \caption*{(b) Achieved MSE}}
  \label{fig:h2}
  \end{subfigure}
\caption{(a) Ergodic sum SE and (b) Achieved MSE with $N = 8, K = 4, M = 4$ and $T_{\mathrm{mse}}=-11$ dB.}
\label{ergodic-rate-SNR}
\end{figure} 

\begin{figure} \centering
 \begin{subfigure}
 \centering{
 \includegraphics[width=0.9\columnwidth]{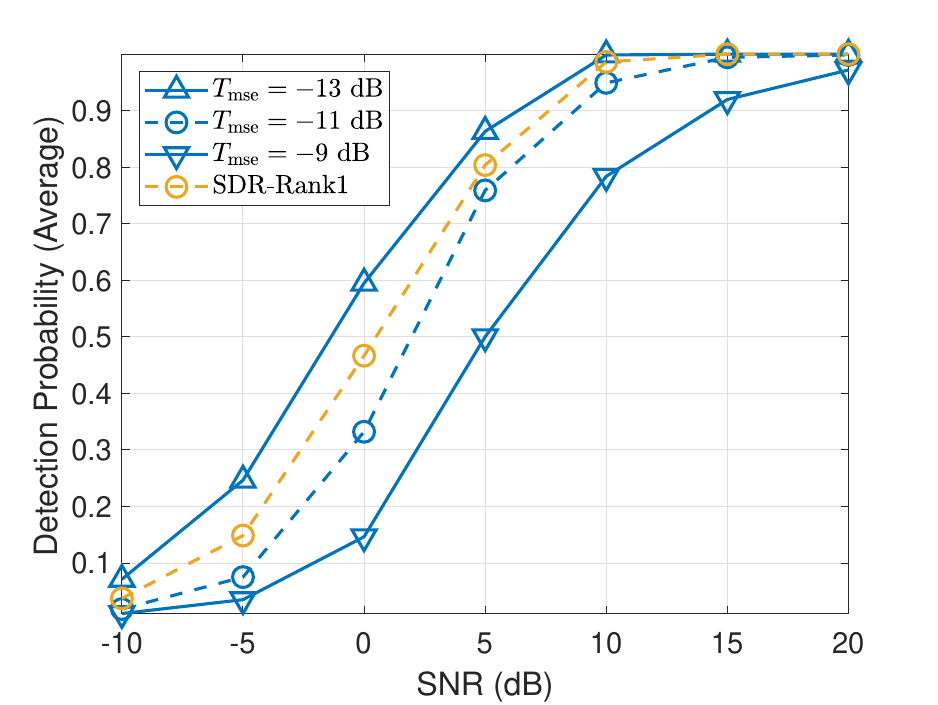}
  \caption*{(a) Average detection probability per SNR (=$P/\sigma^2$)} }
  \end{subfigure}
  \begin{subfigure}
  \centering{
  \includegraphics[width=0.9\columnwidth]
  {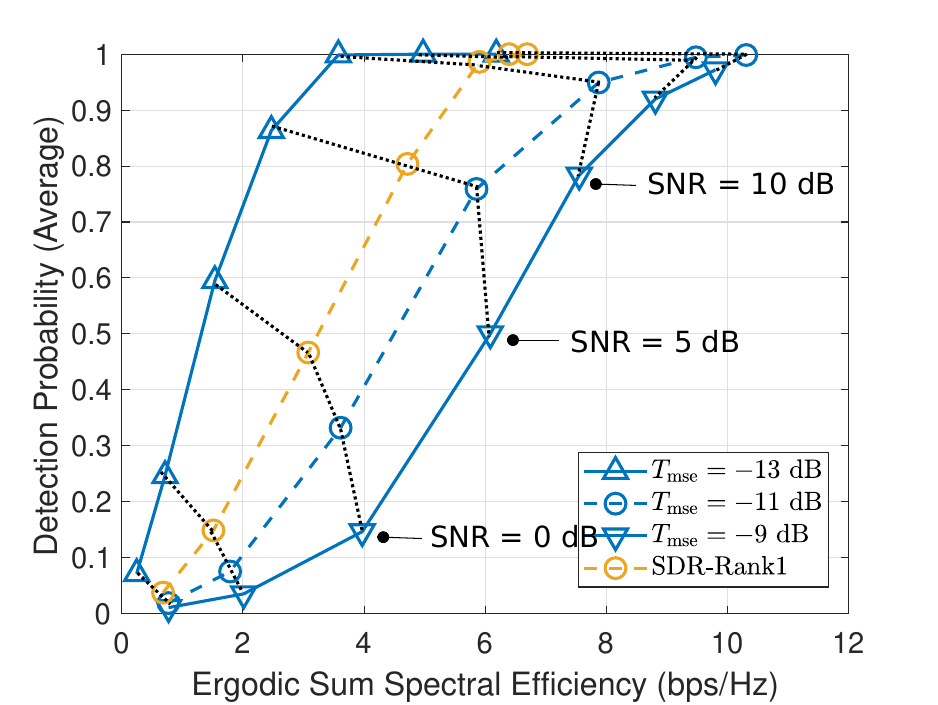} }
 \caption*{(b) Average detection probability vs. ergodic sum SE} 
  \end{subfigure}
\caption{Average detection probability with $N = 8, K = 4, M = 4,$ and $\eta_{k, \ell} = 0.9 \ \forall (k, \ell)$.}
\label{fig:det}
\end{figure}

\subsection{Performance per SNR}
In Fig.~\ref{ergodic-rate-SNR}-(a), we present the ergodic sum spectral efficiency (SE) versus SNR for the proposed RSMA-based ISAC precoding and baseline methods. The SDR-based algorithm without rank-1 approximation achieves identical SE to GPI-ISAC (no RS) but with lower MSE, as shown in Fig.~\ref{ergodic-rate-SNR}-(b). However, this approach is impractical, as feasible beamforming vectors require rank-1 approximation. Consequently, GPI-ISAC (no RS) outperforms the SDR-based algorithm with rank-1 approximation (SDR-Rank1) and regularized zero-forcing (RZF) in SE for \( T_{\text{mse}} = -11 \) dB, while closely satisfying the MSE constraint, as depicted in Fig.~\ref{ergodic-rate-SNR}-(b). This performance gap arises because SDR-Rank1 prioritizes radar-oriented beamforming, limiting communication efficiency. \textcolor{black}{In contrast, the proposed RSMA-based ISAC precoding achieves significant SE gains over GPI-ISAC (no RS) while maintaining equivalent MSE. RSMA’s advantage stems from its common stream, which mitigates multi-user and sensing-to-communication interference in zero-feedback scenarios with imperfect CSIT. By encoding common message parts into a shared stream, decodable by all users via SIC, RSMA aligns with radar beampattern requirements, reducing interference from sensing signals, as explained by \cite{xu:jstsp:22}. This robustness, detailed in \cite{RSMA-ten-promising, Park:2023}, ensures superior SE under CSIT inaccuracies as evidenced in Fig.~\ref{ergodic-rate-SNR}-(a).}

\subsection{Detection probability}
In Fig. \ref{fig:det}-(a), we plot the detection probability as a function of the transmit SNR as studied in \cite{detection-prob}. In particular, the range from -13 to 9 dB corresponds to the achievable region of our proposed method determined by adjustment of \(T_{\mathrm{mse}}\). When the MSE of the target is lower, a more accurate sensing beam is made, leading to a higher detection probability. Specifically, for \(T_{\mathrm{mse}} = -11\) dB, this region is considered suitable for ISAC, where sensing and communication are performed simultaneously. In this case, we compared the performance of the proposed method with the SDR-Rank 1 approach. The SDR-based method, which solves \eqref{SDR-obj}-\eqref{SDR-const} appears to improve the detection performance under the same \(T_{\mathrm{mse}}\). To investigate the trade-off in terms of SE, Fig. \ref{fig:det}-(b) illustrates the ergodic sum SE corresponding to each point in Fig. \ref{fig:det}-(a). Points connected by dashed lines represent the same transmit SNR. As expected, the SDR approach achieves better sensing performance, but falls short in SE compared to the proposed method. Interestingly, this trade-off relationship is evident up to an SNR of 0 dB. However, in higher SNR regions, the proposed method exhibits only marginally lower detection probability compared to the SDR-based approach, while offering substantial SE gains. This highlights the superiority of the proposed method in practical ISAC scenarios.

\begin{figure}[t]
\centering
\includegraphics[width=0.85\columnwidth]{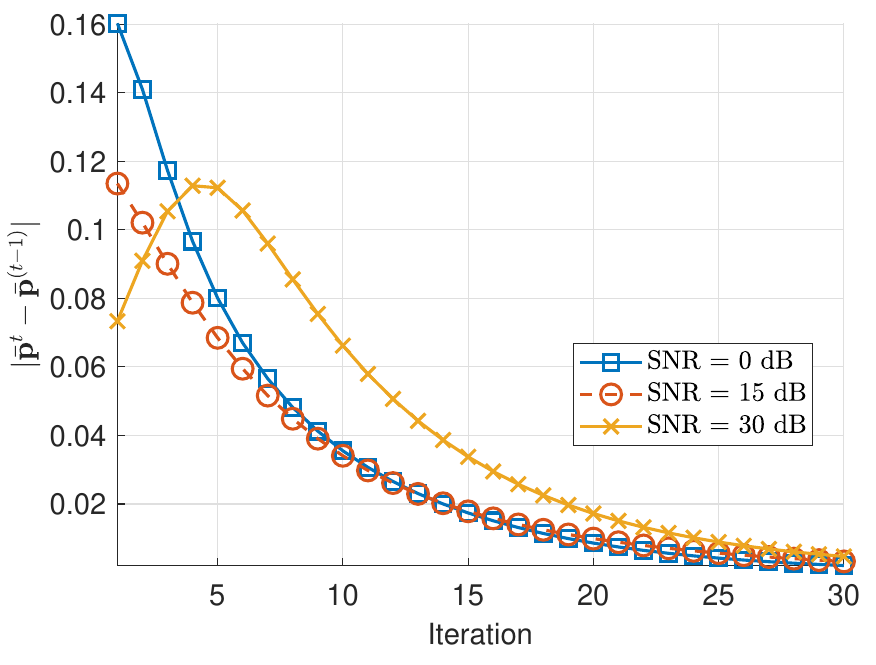}
\caption{Convergence behavior of the proposed GPI-ISAC-RS algorithm over iterations, showing ergodic sum SE versus iteration count for various channel realizations.}
\label{fig:convergence}
\end{figure}

{\color{black}{
\subsection{Convergence}
In Fig. \ref{fig:convergence}, we illustrate the numerical convergence behavior of the proposed framework. As shown in the figure, the algorithm generally converges within approximately $15-20$ iterations, while higher SNR levels require more iterations. We note that, as well observed in \cite{Park:2023}, higher SNR levels demand more delicate adjustments to the precoder due to the increased amount of interference, which delays convergence.
}}

\textcolor{black}{\begin{remark}(Computational complexity analysis)
\normalfont The proposed GPI-ISAC algorithm optimizes precoding with a complexity driven by the inversion of the block diagonal matrix ${\bf{\Omega}}(\bar {\bf{p}})$, comprising $(K+M)$ submatrices of size $N \times N$, requiring $\mathcal{O}(N^3(K+M))$ per iteration. Including the feasibility check, the total complexity is $\mathcal{O}(\xi_{\rm out}(LN(K+M) + \xi_{\rm in}N^3(K+M)))$, where $\xi_{\rm in}$ and $\xi_{\rm out}$ are inner and outer iterations, and $L$ is the number of constraints. For typical cases where $L \leq \xi_{\rm in}N^2$, this simplifies to $\mathcal{O}(\xi_{\rm out}\xi_{\rm in}N^3(K+M))$. In contrast, the QSDP-based approach in \cite{liu:tsp:20, liu:tsp:18} incurs $\mathcal{O}(K^{6.5}M^{6.5}\log(1/\epsilon))$, which scales poorly for large $M$. For massive MIMO-ISAC systems with numerous antennas, our method’s linear scaling in $K+M$ offers significant computational savings, crucial for low-latency applications.
\end{remark}}

\begin{figure} \centering
 \begin{subfigure}
 \centering{
 \includegraphics[width=0.9\columnwidth]{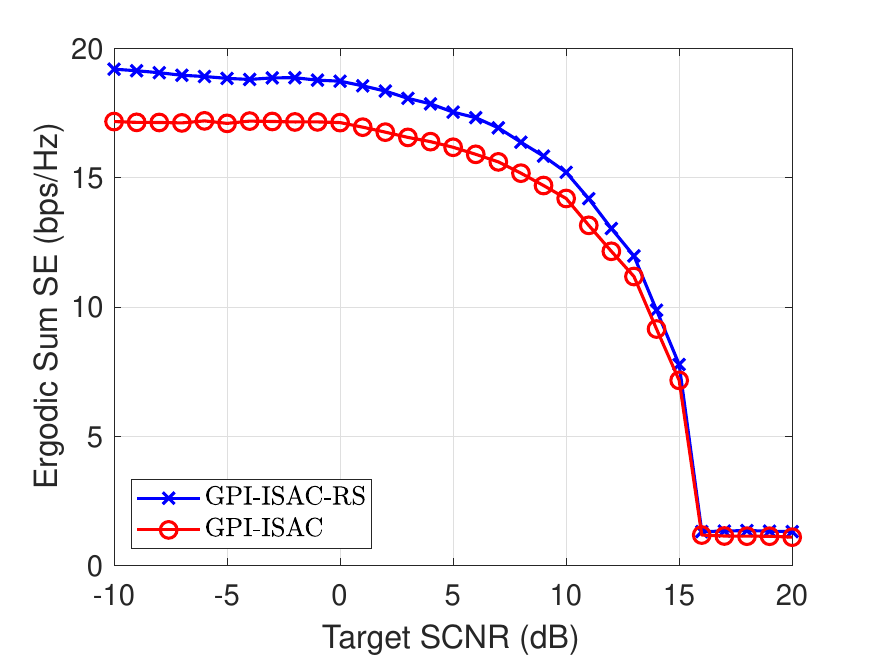}
  \caption*{(a) Ergodic sum SE per target SCNR} }
  \end{subfigure}
  \begin{subfigure}
  \centering{
  \includegraphics[width=0.9\columnwidth]
  {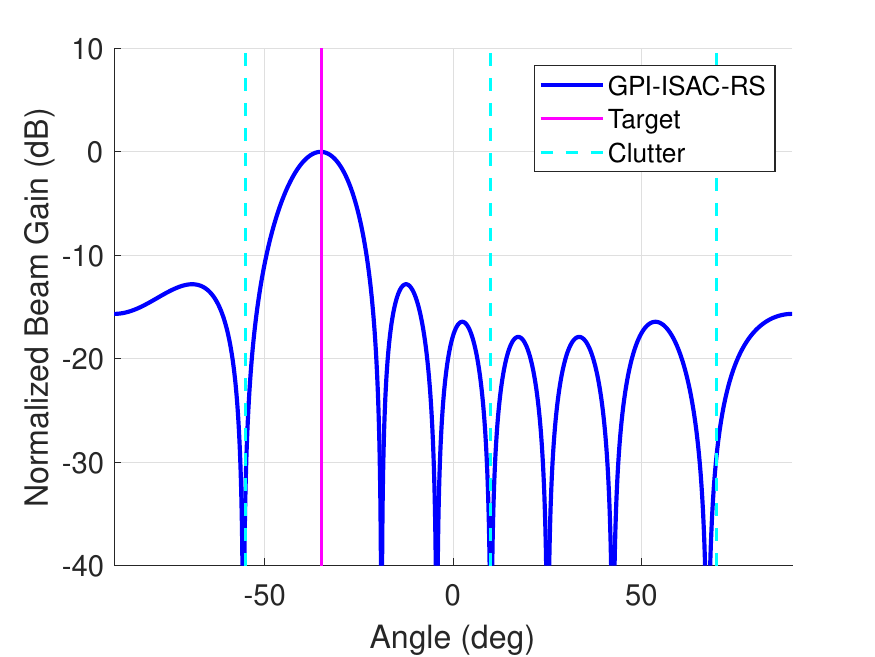} }
 \caption*{(b) Sensing beam pattern with target SCNR = 15 dB} 
  \end{subfigure}
\caption{Ergodic sum SE and sensing beam pattern when considering SCNR as a new sensing metric. $N = 8, K = 4, M = 4,$ and $\eta_{k, \ell} = 0.9 \ \forall (k, \ell)$.}
\label{fig:scnr}
\end{figure}

{\color{black}{
\subsection{Different sensing metrics: SCNR}
Even that the proposed GPI-ISAC-RS mainly considers the beam-pattern MSE as a sensing performance metric, it is also possible to extend it by incorporating different sensing metrics. 
We demonstrate that our method is also applicable when SCNR is used as a sensing metric. 
In sensing, a direct line-of-sight is often required, and signals reflected from non-target objects, termed clutter, degrade sensing accuracy \cite{choi:twc:24}. The received radar echo signal is modeled as:
${\bf y}_{\rm R} = {\bf G}_{\rm tar}{\bf x} + {\bf G}_{\rm cl}{\bf x} + {\bf n}_{\rm R}$,    
where \({\bf G}_{\rm tar} = \sum_{i=1}^S \beta_i^{\rm tar} {\bf a}_{\rm r}(\theta^{\rm tar}_i){\bf a}_{\rm t}^{\sf H}(\theta^{\rm tar}_i)\) represents target reflections, \({\bf G}_{\rm cl} = \sum_{j=1}^C \beta_j^{\rm cl} {\bf a}_{\rm r}(\theta^{\rm cl}_j){\bf a}_{\rm t}^{\sf H}(\theta^{\rm cl}_j)\) denotes clutter, and \({\bf n}_{\rm R} \sim \mathcal{CN}({\bf 0}, \sigma^2_{\rm R}{\bf I}_N)\) is noise. Here, \(S\) and \(C\) are the number of targets and clutters, \(\beta_i^{\rm tar}\), \(\beta_j^{\rm cl}\) are reflection coefficients, and \(\theta_i^{\rm tar}\), \(\theta_j^{\rm cl}\) are directions, with corresponding array response vectors \({\bf a}_{\rm t}(\theta)\), \({\bf a}_{\rm r}(\theta)\). 
Based on this, the SCNR is derived as 
\begin{align}
\gamma_{\rm R}(\bar {\bf{p}}) 
 &= \frac{\bar {\bf p}^{\sf H}({\bf I}_{K+M}\otimes{\bf G}_{\rm tar})^{\sf H}({\bf I}_{K+M}\otimes{\bf G}_{\rm tar})\bar {\bf p}}{\bar {\bf p}^{\sf H}({\bf I}_{K+M}\otimes{\bf G}_{\rm cl})^{\sf H}({\bf I}_{K+M}\otimes{\bf G}_{\rm cl})\bar {\bf p} + \frac{N}{P}\sigma^2_{\rm R}} \label{eq:rev:scnr} \\   
&= \frac{\bar {\bf p}^{\sf H}\bar {\bf{G}}_{\rm tar}\bar {\bf p}}{\bar {\bf p}^{\sf H} \bar {\bf{G}}_{\rm cl}\bar {\bf p} },
\end{align}
which effectively incorporates clutter impact, making it suitable for radar tracking with known target and clutter positions. 
Based on this, the SCNR-constrained ISAC optimization problem is formulated based on $\mathscr{P}_{2}$ as 
\begin{align}
    \mathscr{P}_{\text{scnr}}: & \mathop \textrm{maximize}_{\bar{\mathbf{p}}} \ \bar{R}^{\textrm{LSE}}_{\mathrm{c}}
(\bar{\mathbf{p}}) + \sum_{k=1}^{K}\bar{R}_{k}(\bar{\mathbf{p}})\label{P2_scnr}\\
    &\textrm{subject to } \frac{\bar {\bf p}^{\sf H} \bar {\bf G}_{\rm tar}\bar {\bf p}}{\bar {\bf p}^{\sf H}\bar {\bf G}_{\rm cl}\bar {\bf p}} \ge T_{\rm scnr}.\label{P2-constraint_scnr}
\end{align}
To solve this, we adopt the similar approach to the proposed GPI-ISAC-RS. We obtain the first-order KKT condition of the problem $\mathscr{P}_{\text{scnr}}$ as 
${\bf \Upsilon}(\bar {\bf p}) \bar {\bf p} = {\bf \Xi}(\bar {\bf p}) \eta(\bar {\bf p}) \bar {\bf p}$.  
Here, ${\bf \Upsilon}(\bar {\bf p})$ and ${\bf \Xi}(\bar {\bf p})$ are obtained by replacing the sensing-related term in $\mathbf{L}(\bar{\mathbf{p}}, \mu)$ \eqref{L(p)} and $\mathbf{R}(\bar{\mathbf{p}}, \mu)$ \eqref{R(p)} by
\begin{align}
     \mu \frac{\bar {\bf G}_{\rm tar}}{\bar {\bf p}^{\sf H} \bar {\bf G}_{\rm tar} \bar {\bf p}}, \; \mu \frac{\bar {\bf G}_{\rm cl}}{\bar {\bf p}^{\sf H} \bar {\bf G}_{\rm cl} \bar {\bf p}},
\end{align}
respectively, where $\mu$ indicates the Lagrangian multiplier. 
Building on this, we find the principal eigenvector that solves the KKT condition by slightly modifying the GPI-ISAC-RS. Due to space limitations, we omit the detailed development. 

In Fig.~\ref{fig:scnr}-(a), we observe that the proposed GPI-ISAC-RS consistently outperforms the method in \cite{choi:twc:24} in all SCNR targets. 
Clearly, these gains come from the use of RSMA combined with the proposed ECM estimation, which effectively suppresses the interference caused by DL CSI reconstruction without no direct feedback. 
It is notable that the sensing beam pattern is suitably formed by incorporating the target and clutter directions as shown in Fig.~\ref{fig:scnr}-(b). 
}}

\begin{figure} \centering
 \begin{subfigure}
 \centering{
 \includegraphics[width=0.9\columnwidth]{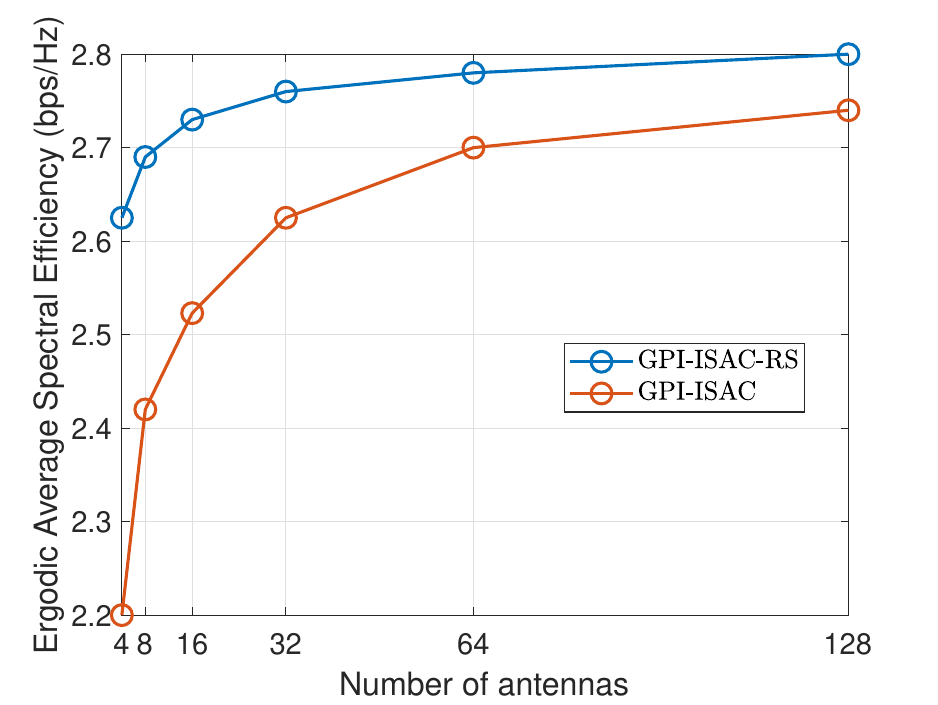}
  \caption*{(a) Ergodic sum SE per number of antennas} }
  \end{subfigure}
  \begin{subfigure}
  \centering{
  \includegraphics[width=0.9\columnwidth]
  {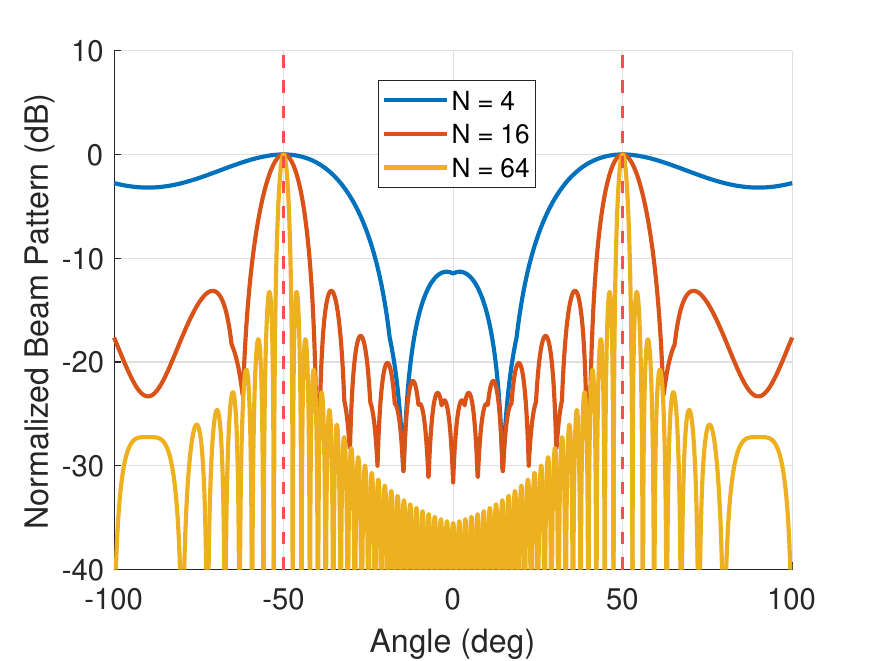} }
 \caption*{(b) Sensing beam pattern with target MSE = -9 dB. The targets are located at -50\degree and 50\degree.} 
  \end{subfigure}
\caption{Ergodic sum SE and sensing beam pattern in large scale antenna regimes. $K = N/2$ is assumed.  }
\label{fig:largeN}
\end{figure}

{\color{black}{
\subsection{Large scale antenna regime}
We compare the performance of the proposed method with that of previous ISAC precoding method \cite{choi:twc:24} as increasing the number of transmit antennas. We fix the number of users to be half the number of antennas (i.e., $K = N/2$), and plot the average SE per user (instead of the sum SE). 
As shown in Fig.~\ref{fig:largeN}-(a), the proposed method consistently outperforms GPI-ISAC across all antenna configurations. The relative gain becomes less pronounced as the number of antennas increases. This is because as $N$ grows, the channel vectors become more orthogonal \cite{kim:arxiv:24}, thereby reducing the interference. This diminishes the relative benefit of the proposed precoding. 
In Fig.~\ref{fig:largeN}-(b), as $N$ increases, the sensing beam becomes sharper at the target direction while the gain at non-target directions is well suppressed. This confirms that the proposed method effectively exploits the increased spatial DoF in large scale antenna regime to enhance radar resolution. 
}}

\section{Conclusion}
This paper presents a precoding method for FDD MIMO ISAC systems that removes the need for channel feedback, thereby simultaneously supporting sensing and communication functionalities with low-latency characteristics. In particular, we aim to maximize the ergodic sum spectral efficiency while satisfying a beam pattern constraint based on the MSE. By reconstructing the DL channel solely from UL training data, we reduce feedback overhead and compensate for CSI imperfections by estimating the ECM. In addition, to mitigate interference between sensing and communication functionality, we consider RSMA that is known to provide robustness toward both objectives. Subsequently, we optimize the RSMA precoder using the KKT conditions to jointly update the precoding vector and Lagrange multipliers, wherein the principle of NEPv is used to resolve the non-convexity. Our extensive simulation results demonstrate that our method offers precise and versatile beam pattern control and substantial SE gains, outperforming current techniques in the sensing-communication performances. 

\section*{Appendix A}
\section*{Proof of Lemma 1}
We define the Lagrangian function as
\begin{align}
    \mathcal{L}(\bar{\mathbf{p}}, \nu) &= -\frac{1}{\eta}\log\left(\frac{1}{K}\sum_{k=1}^{K}\exp\left(-\eta\log_2\left(\frac{\bar{\mathbf{p}}^{\sf{H}}\mathbf{U}_\mathrm{c}(k)\bar{\mathbf{p}}}{\bar{\mathbf{p}}^{\sf H}\mathbf{V}_{\mathrm{c}}(k)\bar{\mathbf{p}}}\right)\right)\right) \nonumber\\
    &\quad + \sum_{k=1}^{K}\log_2\left(\frac{\bar{\mathbf{p}}^{\sf{H}}\mathbf{U}_{k}\bar{\mathbf{p}}}{\bar{\mathbf{p}}^{\sf{H}}\mathbf{V}_{k}\bar{\mathbf{p}}}\right) \nonumber\\
    &\quad + \nu\left(\frac{1}{LP^2}\sum_{\ell=1}^{L}\left|\bar{\mathbf{p}}^{\sf{H}}\mathbf{A}(\theta_\ell)\bar{\mathbf{p}} - \bar{\mathbf{p}}^{\sf{H}}\mathbf{T}(\theta_\ell)\bar{\mathbf{p}}\right|^2-\frac{T_{\text{mse}}}{P^2}\right) \\
    &\triangleq \log_2(\zeta(\bar{\mathbf{p}}, \nu)).
    \label{Lagrangian}
\end{align}
By differentiating \eqref{Lagrangian}, we find that
\begin{align}
    \frac{\partial \mathcal{L}(\bar{\mathbf{p}},\nu)}{{\partial\bar{\mathbf{p}}}}=\frac{\mathcal{L}(\bar{\mathbf{p}},\nu)}{\partial\zeta(\bar{\mathbf{p}}, \nu)}\frac{\partial\zeta(\bar{\mathbf{p}}, \nu)}{\partial\bar{\mathbf{p}}}=\frac{1}{\zeta(\bar{\mathbf{p}}, \nu)\log2}\frac{\partial \zeta(\bar{\mathbf{p}}, \nu)}{\partial\bar{\mathbf{p}}}. \label{temp}
\end{align}
For the stationarity condition, our aim is to find a precoder that satisfies $\partial \mathcal{L}(\bar{\mathbf{p}}, \nu)/\partial\bar{\mathbf{p}}=0$, or equivalently \eqref{temp} equal to zero. Accordingly, $\partial\zeta(\bar{\mathbf{p}}, \nu)/{\partial\bar{\mathbf{p}}}$ is given by
\begin{align}
    &\frac{\partial\zeta(\bar{\mathbf{p}}, \nu)}{\partial\bar{\mathbf{p}}}=\zeta(\bar{\mathbf{p}}, \nu)\Bigg[\sum_{k=1}^{K}\left\{\frac{\mathrm{exp}\left(-\eta\frac{\bar{\mathbf{p}}^{\sf{H}}\mathbf{U}_{\mathrm{c}}(k)\bar{\mathbf{p}}}{\bar{\mathbf{p}}^{\sf{H}}\mathbf{V}_\mathrm{c}(k)\bar{\mathbf{p}}}\right)}{\sum_{j=1}^{K}\mathrm{exp}\left(-\eta\log_2\frac{\bar{\mathbf{p}}^{\sf{H}}\mathbf{U}_{\mathrm{c}}(j)\bar{\mathbf{p}}}{\bar{\mathbf{p}}^{\sf{H}}\mathbf{V}_{\mathrm{c}}(j)\bar{\mathbf{p}}}\right)}\right\} \nonumber \\
    &\quad\quad\quad\times\left(\frac{\mathbf{U}_{\mathrm{c}}(k)\bar{\mathbf{p}}}{\mathbf{p}^{\sf H}\mathbf{U_{\mathrm{c}}}(k)\bar{\mathbf{p}}}-\frac{\mathbf{V}_{\mathrm{c}}(k)\bar{\mathbf{p}}}{\mathbf{p}^{\sf H}\mathbf{V_{\mathrm{c}}}(k)\bar{\mathbf{p}}}\right)+\sum_{k=1}^K\left(\frac{\mathbf{U}_k\bar{\mathbf{p}}}{\bar{\mathbf{p}}^{\sf H}\mathbf{U}_{k}\bar{\mathbf{p}}}-\frac{\mathbf{V}_k\bar{\mathbf{p}}}{\bar{\mathbf{p}}^{\sf H}\mathbf{V}_{k}\bar{\mathbf{p}}}\right)\nonumber \\
    &+\frac{4\nu\log2}{LP^2}\sum_{u=1}^{L}\left(\bar{\mathbf{p}}^{\sf H}\mathbf{A}(\mathrm{\theta}_u)\bar{\mathbf{p}}-\bar{\mathbf{p}}^{\sf H}\mathbf{T(\mathrm{\theta}}_u)\bar{\mathbf{p}}\right)\times\left(\mathbf{A}(\theta_u)\bar{\mathbf{p}}-\mathbf{T}(\theta_u)\bar{\mathbf{p}}\right)\Bigg] \label{eq:derv}.
\end{align}
Therefore, equating \eqref{eq:derv} to zero and reformulating it as expressed in \eqref{FONC}, $\mathbf{L}(\bar{\mathbf{p}}, \nu)$, $\mathbf{R}(\bar{\mathbf{p}}, \nu)$ and $\mathbf{\zeta}(\bar{\mathbf{p}}, \nu)$ can be obtained respectively as follows.
\begin{align}
    &\CMcal{L}(\bar{\mathbf{p}}, \nu) =\zeta_{\mathrm{num}}(\bar{\mathbf{p}}, \nu)\times\nonumber\\ 
    &\Bigg[\sum_{k=1}^{K}\left\{\frac{\exp\left(-\eta\frac{\bar{\mathbf{p}}^{\sf{H}}\mathbf{U}_{\mathrm{c}}(k)\bar{\mathbf{p}}}{\bar{\mathbf{p}}^{\sf{H}}\mathbf{V}_\mathrm{c}(k)\bar{\mathbf{p}}}\right)}{\sum_{j=1}^{K}\exp\left(-\eta\log_2\frac{\bar{\mathbf{p}}^{\sf{H}}\mathbf{U}_{\mathrm{c}}(j)\bar{\mathbf{p}}}{\bar{\mathbf{p}}^{\sf{H}}\mathbf{V}_{\mathrm{c}}(j)\bar{\mathbf{p}}}\right)}\right\}\frac{\mathbf{U}_{\mathrm{c}}(k)}{\bar{\mathbf{p}}^{\sf H}\mathbf{U}_{\mathrm{c}}(k)\bar{\mathbf{p}}}+\sum_{k=1}^K\left(\frac{\mathbf{U}_k}{\bar{\mathbf{p}}^{\sf H}\mathbf{U}_{k}\bar{\mathbf{p}}}\right) \nonumber\\
    &\quad +\frac{4\nu\log2}{LP^2}\sum^L_{u=1}\left((\bar{\mathbf{p}}^{\sf H}\mathbf{A}(\mathrm{\theta}_u)\bar{\mathbf{p}})\mathbf{A}(\mathrm{\theta}_u) +(\bar{\mathbf{p}}^{\sf H}\mathbf{T}(\mathrm{\theta}_u)\bar{\mathbf{p}})\mathbf{T}(\mathrm{\theta}_u)\right)\Bigg],
\end{align}

\begin{align}
    &\mathbf{R}(\bar{\mathbf{p}}, \nu) = \zeta_{\mathrm{den}}(\bar{\mathbf{p}}, \nu) \times\nonumber\\ 
    &\Bigg[\sum_{k=1}^{K}\left\{\frac{\exp\left(-\eta\frac{\bar{\mathbf{p}}^{\sf{H}}\mathbf{U}_{\mathrm{c}}(k)\bar{\mathbf{p}}}{\bar{\mathbf{p}}^{\sf{H}}\mathbf{V}_\mathrm{c}(k)\bar{\mathbf{p}}}\right)}{\sum_{j=1}^{K}\exp\left(-\eta\log_2\frac{\bar{\mathbf{p}}^{\sf{H}}\mathbf{U}_{\mathrm{c}}(j)\bar{\mathbf{p}}}{\bar{\mathbf{p}}^{\sf{H}}\mathbf{V}_{\mathrm{c}}(j)\bar{\mathbf{p}}}\right)}\right\}\frac{\mathbf{V}_{\mathrm{c}}(k)}{\bar{\mathbf{p}}^{\sf H}\mathbf{V}_{\mathrm{c}}(k)\bar{\mathbf{p}}}\nonumber+\sum_{k=1}^K\left(\frac{\mathbf{V}_k}{\bar{\mathbf{p}}^{\sf H}\mathbf{V}_{k}\bar{\mathbf{p}}}\right) \nonumber\\
    &\quad +\frac{4\nu\log2}{LP^2}\sum^L_{u=1}\left((\bar{\mathbf{p}}^{\sf H}\mathbf{A}(\mathrm{\theta}_u)\bar{\mathbf{p}})\mathbf{T}(\mathrm{\theta}_u) +(\bar{\mathbf{p}}^{\sf H}\mathbf{T}(\mathrm{\theta}_u)\bar{\mathbf{p}})\mathbf{A}(\mathrm{\theta}_u)\right)\Bigg],
\end{align}
and
\begin{align}
    \zeta(\bar{\mathbf{p}}, \nu)&=\left \{\frac{1}{K}\sum _{k = 1}^{K} {\exp \!\left ({-\eta\log _{2} \left ({\frac {\bar {\mathbf{p}}^{\sf H} \mathbf{U}_{\mathrm{c}}(k)\bar {\mathbf{p}}}{\bar {\mathbf{p}}^{\sf H} \mathbf{V}_{\mathrm{c}}(k)\bar{\mathbf{p}} }}\right) }\right)}\right \}^{-\frac{1} {\eta\log _{2} e}}\nonumber\\ 
    &\quad\times\prod_{k=1}^K\frac{\bar{\mathbf{p}}^{\sf H}\mathbf{U}_k\bar{\mathbf{p}}}{\bar{\mathbf{p}}^{\sf H}\mathbf{V}_k\bar{\mathbf{p}}} \times 2^{\nu\left(\frac{1}{LP^2}\sum_{u=1}^{L}\left|\bar{\mathbf{p}}^{\sf{H}}\mathbf{A}(\theta_u)\bar{\mathbf{p}} - \bar{\mathbf{p}}^{\sf{H}}\mathbf{T}(\theta_u)\bar{\mathbf{p}}\right|^2-\frac{T_{\mathrm{mse}}}{P^2}\right)}\\
    &=\frac{\zeta_{\mathrm{num}}(\bar{\mathbf{p}}, \nu)}{\zeta_{\mathrm{den}}(\bar{\mathbf{p}}, \nu)}.
\end{align}
This completes the proof.
\bibliographystyle{IEEEtran}
\bibliography{reference}

\begin{thebibliography}{10}
\providecommand{\url}[1]{#1}
\csname url@samestyle\endcsname
\providecommand{\newblock}{\relax}
\providecommand{\bibinfo}[2]{#2}
\providecommand{\BIBentrySTDinterwordspacing}{\spaceskip=0pt\relax}
\providecommand{\BIBentryALTinterwordstretchfactor}{4}
\providecommand{\BIBentryALTinterwordspacing}{\spaceskip=\fontdimen2\font plus
\BIBentryALTinterwordstretchfactor\fontdimen3\font minus
  \fontdimen4\font\relax}
\providecommand{\BIBforeignlanguage}[2]{{%
\expandafter\ifx\csname l@#1\endcsname\relax
\typeout{** WARNING: IEEEtran.bst: No hyphenation pattern has been}%
\typeout{** loaded for the language `#1'. Using the pattern for}%
\typeout{** the default language instead.}%
\else
\language=\csname l@#1\endcsname
\fi
#2}}
\providecommand{\BIBdecl}{\relax}
\BIBdecl

\bibitem{liu:survery:22}
A.~Liu \emph{et~al.}, ``A survey on fundamental limits of integrated sensing
  and communication,'' \emph{IEEE Commun. Surv. Tutor.}, vol.~24, no.~2, pp.
  994--1034, 2022.

\bibitem{choi:twc:24}
J.~Choi, J.~Park, N.~Lee, and A.~Alkhateeb, ``Joint and robust beamforming
  framework for integrated sensing and communication systems,'' \emph{IEEE
  Trans. Wireless Commun.}, vol.~23, no.~11, pp. 17\,602--17\,618, 2024.

\bibitem{liu:tsp:20}
X.~Liu, T.~Huang, N.~Shlezinger, Y.~Liu, J.~Zhou, and Y.~C. Eldar, ``Joint
  transmit beamforming for multiuser {MIMO} communications and {MIMO} radar,''
  \emph{IEEE Trans. Signal Process.}, vol.~68, pp. 3929--3944, 2020.

\bibitem{xu:jstsp:22}
C.~Xu, B.~Clerckx, S.~Chen, Y.~Mao, and J.~Zhang, ``Rate-splitting multiple
  access for multi-antenna joint radar and communications,'' \emph{IEEE J. Sel.
  Topics Signal Process.}, vol.~15, no.~6, pp. 1332--1347, 2021.

\bibitem{wang:twc:24}
Z.~Wang, X.~Mu, and Y.~Liu, ``Bidirectional integrated sensing and
  communication: Full-duplex or half-duplex?'' \emph{IEEE Trans. Wireless
  Commun.}, vol.~23, no.~8, pp. 8184--8199, Aug. 2024.

\bibitem{liu:arxiv:22}
\BIBentryALTinterwordspacing
L.~Liu, S.~Zhang, R.~Du, T.~X. Han, and S.~Cui, ``Networked sensing in {6G}
  cellular networks: Opportunities and challenges,'' \emph{ArXiv Preprint},
  2022. [Online]. Available: \url{https://arxiv.org/abs/2206.00493}
\BIBentrySTDinterwordspacing

\bibitem{FDD-caire-2023}
M.~B. Khalilsarai, Y.~Song, T.~Yang, and G.~Caire, ``{FDD} massive {MIMO}
  channel training: Optimal rate-distortion bounds and the spectral efficiency
  of {“One-Shot”} schemes,'' \emph{IEEE Trans. Wireless Commun.}, vol.~22,
  no.~9, pp. 6018--6032, 2023.

\bibitem{xu:access:14}
Y.~Xu, G.~Yue, and S.~Mao, ``User grouping for massive {MIMO} in {FDD} systems:
  New design methods and analysis,'' \emph{IEEE Access}, vol.~2, pp. 947--959,
  2014.

\bibitem{liu:tsp:18}
F.~Liu, L.~Zhou, C.~Masouros, A.~Li, W.~Luo, and A.~Petropulu, ``Toward
  dual-functional radar-communication systems: Optimal waveform design,''
  \emph{IEEE Trans. Signal Process.}, vol.~66, no.~16, pp. 4264--4279, Aug.
  2018.

\bibitem{chen:tsp:21}
L.~Chen, F.~Liu, W.~Wang, and C.~Masouros, ``Joint radar-communication
  transmission: A generalized {Pareto} optimization framework,'' \emph{IEEE
  Trans. Signal Process.}, vol.~69, pp. 2752--2765, 2021.

\bibitem{chowdary2024hybrid}
A.~Chowdary, A.~Bazzi, and M.~Chafii, ``On hybrid radar fusion for integrated
  sensing and communication,'' \emph{IEEE Trans. Wireless Commun.}, vol.~23,
  no.~8, pp. 8984--9000, 2024.

\bibitem{caire:tit:10}
G.~Caire, N.~Jindal, M.~Kobayashi, and N.~Ravindran, ``Multiuser {MIMO}
  achievable rates with downlink training and channel state feedback,''
  \emph{IEEE Trans. Inf. Theory}, vol.~56, no.~6, pp. 2845--2866, 2010.

\bibitem{park:twc:16}
J.~Park, N.~Lee, J.~G. Andrews, and R.~W. Heath~Jr., ``On the optimal feedback
  rate in interference-limited multi-antenna cellular systems,'' \emph{IEEE
  Trans. Wireless Commun.}, vol.~15, no.~8, pp. 5748--5762, 2016.

\bibitem{TD-ISAC}
Q.~Zhang, H.~Sun, X.~Gao, X.~Wang, and Z.~Feng, ``Time-division {ISAC} enabled
  connected automated vehicles cooperation algorithm design and performance
  evaluation,'' \emph{IEEE J. Sel. Areas Commun.}, vol.~40, no.~7, pp.
  2206--2218, 2022.

\bibitem{wang:tsp:24}
S.~Wang, W.~Dai, H.~Wang, and G.~Y. Li, ``Robust waveform design for integrated
  sensing and communication,'' \emph{IEEE Trans. Signal Process.}, vol.~72, pp.
  3122--3138, 2024.

\bibitem{ren:tcom:22}
Z.~Ren, L.~Qiu, J.~Xu, and D.~W.~K. Ng, ``Robust transmit beamforming for
  secure integrated sensing and communication,'' \emph{IEEE J. Sel. Topics
  Signal Process.}, vol.~71, no.~9, pp. 5549--5564, 2023.

\bibitem{luan:tits:22}
M.~Luan, B.~Wang, Z.~Chang, T.~Hämäläinen, and F.~Hu, ``Robust beamforming
  design for {RIS}-aided integrated sensing and communication system,''
  \emph{IEEE Trans. Intell. Transp. Syst.}, vol.~24, no.~6, pp. 6227--6243,
  2023.

\bibitem{bazzi:arxiv:23}
\BIBentryALTinterwordspacing
B.~Ahmad and M.~Chafii, ``Robust integrated sensing and communication
  beamforming for dual-functional radar and communications: Method and
  insights,'' \emph{ArXiv Preprint}, 2023. [Online]. Available:
  \url{https://arxiv.org/abs/2303.07652}
\BIBentrySTDinterwordspacing

\bibitem{Dina}
D.~Vasisht, S.~Kumar, H.~Rahul, and D.~Katabi, ``Eliminating channel feedback
  in next-generation cellular networks,'' in \emph{Proc. of the ACM SIGCOMM
  Conf.}, 2016, pp. 398--411.

\bibitem{T.Choi:2021}
T.~Choi, F.~Rottenberg, J.~Gómez-Ponce, A.~Ramesh, P.~Luo, C.~J. Zhang, and
  A.~F. Molisch, ``Experimental investigation of frequency domain channel
  extrapolation in massive {MIMO} systems for zero-feedback {FDD},'' \emph{IEEE
  Trans. Wireless Commun.}, vol.~20, no.~1, pp. 710--725, 2021.

\bibitem{Han:2019}
Y.~Han, T.~H. Hsu, C.~K. Wen, K.~K. Wong, and S.~Jin, ``Efficient downlink
  channel reconstruction for {FDD} multi-antenna systems,'' \emph{IEEE Trans.
  Wireless Commun.}, vol.~18, no.~6, pp. 3161--3176, June 2019.

\bibitem{Rottenberg:2020}
F.~Rottenberg, T.~Choi, P.~Luo, C.~J. Zhang, and A.~F. Molisch, ``Performance
  analysis of channel extrapolation in {FDD} massive {MIMO} systems,''
  \emph{IEEE Trans. Wireless Commun.}, vol.~19, no.~4, pp. 2728--2741, Apr.
  2020.

\bibitem{NOMP}
B.~Mamandipoor, D.~Ramasamy, and U.~Madhow, ``Newtonized orthogonal matching
  pursuit: Frequency estimation over the continuum,'' \emph{IEEE Trans. Signal
  Process.}, vol.~64, no.~19, pp. 5066--5081, Oct. 2016.

\bibitem{kim:arxiv:24}
N.~Kim, I.~P. Roberts, and J.~Park, ``Splitting messages in the
  dark-rate-splitting multiple access for {FDD} massive {MIMO} without {CSI}
  feedback,'' \emph{IEEE Trans. Wireless Commun.}, vol.~24, no.~4, pp.
  3320--3332, 2025.

\bibitem{Park:2023}
J.~Park, J.~Choi, N.~Lee, W.~Shin, and H.~V. Poor, ``Rate-splitting multiple
  access for downlink {MIMO}: A generalized power iteration approach,''
  \emph{IEEE Trans. Wireless Commun.}, vol.~22, no.~3, pp. 1588--1603, Mar.
  2023.

\bibitem{Deokhwan}
D.~Han, J.~Park, and N.~Lee, ``{FDD} massive {MIMO} without {CSI} feedback,''
  \emph{IEEE Trans. Wireless Commun.}, vol.~23, no.~5, pp. 4518--4530, 2024.

\bibitem{Zhong:2020}
Z.~Zhong, L.~Fan, and S.~Ge, ``{FDD} massive {MIMO} uplink and downlink channel
  reciprocity properties: Full or partial reciprocity?'' \emph{Proc. IEEE Glob.
  Comm. Conf.}, Dec. 2020.

\bibitem{RSMA-ten-promising}
J.~Park \emph{et~al.}, ``Rate-splitting multiple access for {6G} networks:
  {Ten} promising scenarios and applications,'' \emph{IEEE Network}, vol.~38,
  no.~3, pp. 128--136, 2024.

\bibitem{WMMSE-SAA}
H.~Joudeh and B.~Clerckx, ``Sum-rate maximization for linearly precoded
  downlink multiuser {MISO} systems with partial {CSIT}: A rate-splitting
  approach,'' \emph{IEEE Trans. Commun.}, vol.~64, no.~11, pp. 4847--4861,
  2016.

\bibitem{shakya2024urban}
D.~Shakya, M.~Ying, T.~S. Rappaport, P.~Ma, I.~Al-Wazani, Y.~Wu, Y.~Wang,
  D.~Calin, H.~Poddar, A.~Bazzi \emph{et~al.}, ``Urban outdoor propagation
  measurements and channel models at 6.75 {GHz} {FR1(C)} and 16.95 {GHz} {FR3}
  upper mid-band spectrum for {5G} and {6G},'' \emph{arXiv preprint
  arXiv:2410.17539}, 2024.

\bibitem{GantiLapidoth2000}
A.~Ganti, A.~Lapidoth, and I.~Telatar, ``Mismatched decoding revisited: general
  alphabets, channels with memory, and the wide-band limit,'' \emph{IEEE Trans.
  Inf. Theory}, vol.~46, no.~7, pp. 2315--2328, 2000.

\bibitem{yin-mse:tcl:22}
L.~Yin, Y.~Mao, O.~Dizdar, and B.~Clerckx, ``Rate-splitting multiple access for
  {6G}—part {II}: Interplay with integrated sensing and communications,''
  \emph{IEEE Communications Letters}, vol.~26, no.~10, pp. 2237--2241, 2022.

\bibitem{fan-crb:tsp:22}
F.~Liu, Y.-F. Liu, A.~Li, C.~Masouros, and Y.~C. Eldar, ``{Cramér-Rao} bound
  optimization for joint radar-communication beamforming,'' \emph{IEEE Trans.
  Signal Process.}, vol.~70, pp. 240--253, 2022.

\bibitem{covApprox}
E.~Björnson, L.~Sanguinetti, and M.~Debbah, ``Massive {MIMO} with imperfect
  channel covariance information,'' in \emph{Proc. of Asilomar Conf. on Sign.,
  Syst. and Computers}, 2016, pp. 974--978.

\bibitem{kay1993statistical}
S.~M. Kay, ``Statistical signal processing: estimation theory,'' \emph{Prentice
  Hall}, vol.~1, pp. Chapter--3, 1993.

\bibitem{observedFisher}
B.~G. Lindsay and B.~Li, ``On second-order optimality of the observed {Fisher}
  information,'' \emph{The Annals of Statistics}, vol.~25, no.~5, pp.
  2172--2199, 1997.

\bibitem{efron1978assessing}
B.~Efron and D.~V. Hinkley, ``Assessing the accuracy of the maximum likelihood
  estimator: Observed versus expected {Fisher} information,''
  \emph{Biometrika}, vol.~65, no.~3, pp. 457--483, 1978.

\bibitem{detection-prob}
A.~De~Maio, S.~De~Nicola, Y.~Huang, S.~Zhang, and A.~Farina, ``Code design to
  optimize radar detection performance under accuracy and similarity
  constraints,'' \emph{IEEE Trans. Signal Process.}, vol.~56, no.~11, pp.
  5618--5629, 2008.

\end{thebibliography}
\end{document}